\documentclass[12pt,epsf]{article}

\usepackage{times}
\usepackage{graphicx}
\usepackage{amsfonts}
\usepackage{amsmath}
\usepackage{amssymb}
\usepackage{amstext}
\usepackage{amsthm}

\usepackage{epsfig}
\usepackage{color}
\psfigdriver{dvips}
\usepackage{latexsym}
\usepackage[matrix,curve]{xypic}
\usepackage{rotating}
\rotdriver{dvips}

\begin{document}

\baselineskip 6mm
\renewcommand{\thefootnote}{\fnsymbol{footnote}}

%------------ Hyun Seok's macro's, etc  -----------

\newcommand{\nc}{\newcommand}
\newcommand{\rnc}{\renewcommand}

%\headheight=0truein
%\headsep=0truein
%\topmargin=0truein
%\oddsidemargin=0truein
%\evensidemargin=0truein
%\textheight=9truein
%\textwidth=6.5truein

\rnc{\baselinestretch}{1.24}    % 1.5 spacing btwn text lines
\setlength{\jot}{6pt}       % spacing btwn the rows of an eqnarray
\rnc{\arraystretch}{1.24}   % spacing btwn the rows of a non-eqn array

%%%%%%%%%%%%%%%%%%%%%% Equation Numbering %%%%%%%%%%%%%%%%%%%%%%%
\makeatletter
\rnc{\theequation}{\thesection.\arabic{equation}}
\@addtoreset{equation}{section}
\makeatother

%%%%%%%%%%%%%%%%%%%%%%%%%%%%%%%%%%%%%%%%%%%%%%%%%%%%%%%%%%%%%%%%%
%                                                               %
%                NEW COMMANDS AND MACROS                        %
%                                                               %
%%%%%%%%%%%%%%%%%%%%%%%%%%%%%%%%%%%%%%%%%%%%%%%%%%%%%%%%%%%%%%%%%

%%%%% Simplify some frequently used LaTeX commands %%%%%

\nc{\be}{\begin{equation}}

\nc{\ee}{\end{equation}}

\nc{\bea}{\begin{eqnarray}}

\nc{\eea}{\end{eqnarray}}

\nc{\xx}{\nonumber\\}

\nc{\ct}{\cite}

\nc{\la}{\label}

\nc{\eq}[1]{(\ref{#1})}

\nc{\newcaption}[1]{\centerline{\parbox{6in}{\caption{#1}}}}

\nc{\fig}[3]{

\begin{figure}
\centerline{\epsfxsize=#1\epsfbox{#2.eps}}
\newcaption{#3. \label{#2}}
\end{figure}
}

%%% Caligraphic letters %%%%

\def\CA{{\cal A}}
\def\CC{{\cal C}}
\def\CD{{\cal D}}
\def\CE{{\cal E}}
\def\CF{{\cal F}}
\def\CG{{\cal G}}
\def\CH{{\cal H}}
\def\CK{{\cal K}}
\def\CL{{\cal L}}
\def\CM{{\cal M}}
\def\CN{{\cal N}}
\def\CO{{\cal O}}
\def\CP{{\cal P}}
\def\CR{{\cal R}}
\def\CS{{\cal S}}
\def\CU{{\cal U}}
\def\CV{{\cal V}}
\def\CW{{\cal W}}
\def\CY{{\cal Y}}
\def\CZ{{\cal Z}}

%%% Double line letters %%%

\def\IB{{\hbox{{\rm I}\kern-.2em\hbox{\rm B}}}}
\def\IC{\,\,{\hbox{{\rm I}\kern-.50em\hbox{\bf C}}}}
\def\ID{{\hbox{{\rm I}\kern-.2em\hbox{\rm D}}}}
\def\IF{{\hbox{{\rm I}\kern-.2em\hbox{\rm F}}}}
\def\IH{{\hbox{{\rm I}\kern-.2em\hbox{\rm H}}}}
\def\IN{{\hbox{{\rm I}\kern-.2em\hbox{\rm N}}}}
\def\IP{{\hbox{{\rm I}\kern-.2em\hbox{\rm P}}}}
\def\IR{{\hbox{{\rm I}\kern-.2em\hbox{\rm R}}}}
\def\IZ{{\hbox{{\rm Z}\kern-.4em\hbox{\rm Z}}}}

%%% Greek letters %%%

\def\a{\alpha}
\def\b{\beta}
\def\d{\delta}
\def\ep{\epsilon}
\def\ga{\gamma}
\def\k{\kappa}
\def\l{\lambda}
\def\s{\sigma}
\def\t{\theta}
\def\w{\omega}
\def\G{\Gamma}

%%%%% Mathematical Symbols

\def\half{\frac{1}{2}}
\def\dint#1#2{\int\limits_{#1}^{#2}}
\def\goto{\rightarrow}
\def\para{\parallel}
\def\brac#1{\langle #1 \rangle}
\def\curl{\nabla\times}
\def\div{\nabla\cdot}
\def\p{\partial}

%%%%% Roman pont in math

\def\Tr{{\rm Tr}\,}
\def\det{{\rm det}}

%%%%% Special Letters

\def\vare{\varepsilon}
\def\zbar{\bar{z}}
\def\wbar{\bar{w}}
\def\what#1{\widehat{#1}}

%%%%% For this paper only

\def\ad{\dot{a}}
\def\bd{\dot{b}}
\def\cd{\dot{c}}
\def\dd{\dot{d}}
\def\so{SO(4)}
\def\bfr{{\bf R}}
\def\bfc{{\bf C}}
\def\bfz{{\bf Z}}

\begin{titlepage}

%---------------- preprint number ---------------

\hfill\parbox{3.7cm} {{\tt arXiv:1503.00712}}

\vspace{15mm}

\begin{center}
%------------------------ title ------------------------
{\Large \bf Emergent Spacetime and Cosmic Inflation}

\vspace{10mm}
%---------------- authors and addresses ----------------

Hyun Seok Yang \footnote{hsyang@gist.ac.kr}
\\[10mm]

{\sl Department of Physics and Photon Science, Gwangju Institute of Science and Technology, \\
Gwangju 61005, Korea}

\end{center}

\thispagestyle{empty}

\vskip1cm

%----------------------- abstract ----------------------

\centerline{\bf ABSTRACT}
\vskip 4mm
\noindent
 
We present a novel background-independent framework for cosmic inflation, starting with a matrix model. 
In this framework, inflation is portrayed as a dynamic process responsible for the generation of both space and time. 
This stands in contrast to conventional inflation, which is characterized as a mere (exponential) expansion of an already 
existing spacetime, driven by the vacuum energy associated with an inflaton field.
We observe that the cosmic inflation is triggered by the condensate of Planck
energy into vacuum and responsible for the dynamical emergence of spacetime.  
The emergent spacetime picture admits a background-independent formulation so that the
inflation is described by a conformal Hamiltonian system which requires neither an inflaton field 
nor an {\it ad hoc} inflation potential.
This implies that the emergent spacetime may incapacitate all the rationales to introduce
the multiverse hypothesis. 
\\

%PACS numbers: 04.60.-m, 11.25.Tq, 04.50.-h


\vspace{1cm}

\today

\vspace{2.5cm}

\textit{Published in the Special Issue \textbf{Mathematical Cosmology}: Universe \textbf{10} (2024) 3, 150.}

\end{titlepage}

\renewcommand{\thefootnote}{\arabic{footnote}}
\setcounter{footnote}{0}

\section{Introduction}

History is a mirror to the future. If we do not learn from the mistakes of history, we are doomed to
repeat them.\footnote{George Santayana (1863-1952).}
In the middle of the 19th century, Maxwell's equations for electromagnetic phenomena
predicted the existence of an absolute speed, $c = 2.998 \times 10^8$ m/sec,
which apparently contradicted the Galilean relativity, a cornerstone on which
the Newtonian model of space and time rested. Since most
physicists, by then, had developed deep trust in the Newtonian model, they concluded that Maxwell's
equations can only hold in a specific reference frame, called the ether. However, by doing so,
they reverted back to the Aristotelian view that Nature specifies an absolute rest frame.
It was Einstein to realize the true implication of this quandary: It was asking us to abolish
Newton's absolute time as well as absolute space. The ether was removed by the Einstein's special
relativity by radically modifying the concept of space and time in the Newtonian dynamics.
Time lost its absolute standing and the notion of absolute simultaneity was physically untenable.
Only the four-dimensional spacetime has an absolute meaning. The new paradigm of spacetime
has completely changed the Newtonian world with dramatic consequences.

The physics of the last century had devoted to the study of two pillars: general relativity and
quantum field theory. And the two cornerstones of modern physics can be merged into beautiful
equations, the so-called Einstein equations given by
\begin{equation}\label{1einstein-eq}
    R_{\mu\nu} - \frac{1}{2} g_{\mu\nu} R = 8 \pi G_N T_{\mu\nu},
\end{equation}
where the right-hand side is the energy-momentum tensor whose contents are described by (quantum)
field theories. Although the groundbreaking theories of relativity and quantum mechanics have utterly
changed the way we think about Nature and the Universe, new open problems have emerged which
have not been resolved yet within the paradigm of the 20th century physics.
For example, a short list of them is the cosmological constant problem, the hierarchy problem,
dark energy, dark matter, cosmic inflation and quantum gravity.
In particular, recent developments in cosmology, particle physics and string theory
have led to a radical proposal that there could be an ensemble of universes that might be
completely disconnected from ours \cite{1carr-book}.
Certainly, it would be perverse to claim that nothing exists
beyond the horizon of our observable universe. The observable universe is one causal patch of a much
larger unobservable universe. However, resorting to the concept of the string landscape or multiverse 
in an attempt to address certain notorious issues in theoretical physics through the anthropic argument 
is a challenging approach \cite{1carr-ellis}. 
``And it's pretty unsatisfactory to use
the multiverse hypothesis to explain only things we don't understand."\footnote{Graham Ross
in {\it Quanta magazine} ``At multiverse impasse, a new theory
of scale" (August 18, 2014) and {\it Wired.com} ``Radical new theory could kill
the multiverse hypothesis."}
Reflecting on history, the current situation strongly echoes the era of the hypothetical luminiferous ether 
in the late 19th century. The historical lesson implies that we may need another turn
of the spacetime picture to defend the integrity of physics.

In physical cosmology, cosmic inflation in the early universe is the exponential expansion of space.
Suppose that spacetime evolution is determined by a single scale factor $a(t)$ and
its Hubble expansion rate $H \equiv \frac{\dot{a}}{a}$ according to the cosmological principle
and driven by the dynamics of a scalar field $\phi$, called the inflaton \cite{1guth,1linde}.
Then the Einstein equation \eq{1einstein-eq} reduces to the Friedmann equation
\begin{equation}\label{1friedmann}
    H^2  = \frac{8 \pi G_N}{3} \Big( \frac{1}{2} \dot{\phi}^2 + V(\phi) \Big).
\end{equation}
The evolution equation of the inflaton in the Friedmann universe is described by
\begin{equation}\label{1inflaton-eq}
    \ddot{\phi} + 3 H \dot{\phi} + \frac{\delta V}{\delta \phi} = 0.
\end{equation}
The Friedmann equation \eq{1friedmann} tells us that in the early universe with $V(\phi) \approx V_0$
and $\dot{\phi} \approx 0$, there was an inflationary epoch of the exponential expansion of space,
i.e., $a(t) \propto e^{Ht}$ where $H =  \sqrt{\frac{8 \pi G_N V_0}{3}}$ is called
the inflationary Hubble constant. In order to successfully fit to data, one finds \cite{liddle}
\begin{equation}\label{1inf-energy}
    V_0 \geq ( 2 \times 10^{15} \mathrm{GeV})^4 \approx (10^{-3} M_{P})^4
\end{equation}
where $M_{P} = 1/ \sqrt{8 \pi G_N}$ is the Planck mass.

Let us critically examine the inflationary scenario. 
According to this scenario \cite{1guth,1linde},
inflation is described by the exponential expansion of the universe
in a supercooled false vacuum state that is a metastable state without any fields or particles
but with a large energy density.
It should be emphasized that the inflation scenario so far has been formulated in the context of
effective field theory coupled to general relativity. Thus, in this scenario, the existence of space
and time is {\it a priori} assumed from the beginning although the evolution of spacetime is determined
by Eq. \eq{1einstein-eq}. In other terms, the inflationary scenario does not delineate the generation or 
creation of spacetime but merely signifies the expansion of preexisting spacetime. 
It does not delve into the dynamic origin of spacetime. 
Nevertheless, there has to be a definite beginning so that quantum gravity era 
cannot be avoided in the past even if inflation takes place \cite{1bgv-infl}. 
This implies that the current inflationary scenario is insufficient 
in describing the initial stage of our universe, 
and it necessitates the incorporation of new physics to explore 
the past boundaries of inflating regions.\footnote{Even though 
our work was motivated by Ref. \cite{1bgv-infl}, it should be pointed out that 
the conclusions in \cite{1bgv-infl} can be avoided in the so-called 
emergent universe scenario \cite{euni-rev} and some models were also presented to cure the instabilities 
of the emergent universe \cite{eunin-rev}.} 
One plausible explanation is the occurrence of a quantum creation 
as a beginning of the universe \cite{1hmv}.

The Friedmann equation \eq{1friedmann} reveals that cosmic inflation is triggered by the potential energy associated with an inflaton, 
whose energy scale is in proximity to the Planck energy. Near the Planck energy, quantum gravity effects become strong 
and the effective field theory description may break down.
If one identifies the slowly varying inflaton field $\phi(t)$
with a particle trajectory $x(t)$ and $\dot{\phi}(t)$ with its velocity $v(t) = \dot{x}(t)$,
the evolution equation \eq{1inflaton-eq} tells us that the frictional force, $3H v (t)$,
resulting from the inflating spacetime, is (almost) balanced with
an external force $F (x) = - \frac{dV}{dx}$, i.e.,
\begin{equation}\label{1friction}
 3H \dot{x} (t) \approx F(x),
\end{equation}
because $\ddot{x} \approx 0$ during inflation. This implies that the cosmic inflation as
a dynamical system corresponds to a non-Hamiltonian system.\footnote{Nonetheless,
the friction term does not lead to dissipative energy production. This fact can be seen by observing
that Eq. \eq{1inflaton-eq} can be derived from the first law of thermodynamics, $dE + p dV = V d\rho
+ (\rho + p) dV =0$, where $\rho + p = \dot{\phi}^2$ and $\dot{\rho} = \Big( \ddot{\phi}
+ \frac{\delta V}{\delta \phi} \Big) \dot{\phi}$.}

Recent advancements in string theory have unveiled a remarkable and radical perspective on the nature of gravity. 
One notable example is the AdS/CFT correspondence, which depicts a surprising scenario wherein $U(N)$ gauge theory 
in lower dimensions defines a nonperturbative formulation of quantum gravity in higher dimensions \cite{1ads-cft}. 
In particular, the AdS/CFT duality
shows a typical example of emergent gravity and emergent space because gravity in higher
dimensions is defined by a gravityless field theory in lower dimensions.
Now, numerous examples from string theory illustrate that spacetime is not fundamental but rather emerges only at large distances, 
constituting a classical approximation \cite{1hopo-rev}. 
Consequently, the governing principle in quantum gravity dictates that space and time are an emergent entity.
Since the emergent spacetime, we believe, is a significant new paradigm for quantum gravity,
we aim to apply the emergent spacetime picture to cosmic inflation.
We will propose a background-independent formulation of the cosmic inflation.\footnote{Here
we refer to the background-independent theory in which any spacetime structure is not {\it a priori}
assumed but defined by the theory.}
This means that we do not assume the prior existence of spacetime but
define a spacetime structure as a solution of an underlying background-independent theory
such as matrix models. The inflation in this picture corresponds to a dynamical process to generate
space and time which is very different from the standard inflation simply describing
an (exponential) expansion of a preexisting spacetime.
It turns out that spacetime is emergent from the Planck energy condensate in vacuum that
generates an extremely large {\it Universe}. Our observable patch within cosmic horizon
is a very tiny part $\sim 10^{-26}$ of the entire spacetime.
Originally the multiverse hypothesis has been motivated
by an attempt to explain the anthropic fine-tuning such as
the cosmological constant problem \cite{1weinberg-87}
and boosted by the chaotic and eternal inflation scenarios \cite{1guth,1linde}
and the string landscape derived from the Kaluza-Klein compactification of
string theory \cite{1slands,1susskind}, which are all based on the traditional spacetime picture.
Since emergent spacetime is radically different
from any previous physical theories, all of which describe what happens in a given spacetime,
the multiverse picture must be reexamined from the standpoint of emergent spacetime.
The cosmic inflation from the emergent spacetime picture will certainly open a new prospect
that may cripple all the rationales to introduce the multiverse hypothesis \cite{1hsy-jpcs12,1essay}.

Given that the concept of the multiverse introduces profound conceptual challenges, compelling us to reconsider 
the very foundations of science \cite{1carr-ellis}, it becomes imperative to carefully contemplate the true nature 
of the multiverse. Is it merely a speculative illusion stemming from an incomplete physics, 
akin to the ether in the late 19th century, or does it hold significant relevance 
even within a more complete theoretical framework?
The main purpose of this paper is to illuminate how the emergent spacetime picture
brings about radical changes of physics, especially, regarding to physical cosmology.
In particular, a background-independent theory such as matrix models provides a concrete realization
of the idea of emergent spacetime which has a sufficiently elegant and explanatory power
to defend the integrity of physics against the multiverse hypothesis \cite{1hsy-jpcs12,1essay}.
The emergent spacetime is a completely new paradigm so that the multiverse debate in physics
circles has to seriously take it into account.

This paper is organized as follows.  In Sec. 2, we compactly review the background-independent
formulation of emergent gravity and emergent spacetime in terms of matrix models \cite{1hsy-jhep09,1hsy-review,1q-emg,1h-stein}.
See also closely related works \cite{hsy-ijmp09,1d-berens,1ncmat-rev,1hec-ver,1ferrari}.  
The background-independent formulation of
emergent gravity crucially relies on the fact that noncommutative (NC) space arises as
a vacuum solution of a large $N$ matrix model in the Coulomb branch and this vacuum on the Coulomb branch
admits a separable Hilbert space as quantum mechanics \cite{ly-jkps}.
The gravitational metric is derived from a nontrivial inner automorphism of
the NC algebra $\mathcal{A}_\theta$, in which the NC nature is essential to realize
the emergent gravity.  
An important point is that the matrix model does not presuppose any
spacetime background on which fundamental processes develop.
Rather the background-independent theory provides a mechanism of spacetime generation
such that any spacetime structure including the flat spacetime arises as a solution
of the theory itself.

In Sec. 3, we note that the Planck energy condensate into vacuum must be a dynamical process. 
We show that the cosmic inflation arises as a solution of a time-dependent matrix model, describing the dynamical process of the vacuum energy condensation. 
It turns out that the cosmic inflation corresponds to the
dynamical mechanism for the instantaneous condensation of vacuum energy to enormously spread
out spacetime. It is remarkable to see that the inflation can be described by time-dependent matrices
only without introducing any inflaton field as well as an {\it ad hoc} inflation potential.
Our work is not the first to address physical cosmology using matrix models.
There have been interesting earlier attempts \cite{1matrix-cos}.
In particular, the cosmic inflation was addressed in very interesting works \cite{1jun-sang}
using the Monte Carlo analysis of the type IIB matrix model in Lorentzian signature
and it was found that three out of nine spatial directions start to expand at some critical time
after which exactly (3+1)-dimensions dynamically become macroscopic.

In Sec. 4, we discuss why the cosmic inflation triggered by the Planck energy condensate 
into vacuum must be a single event \cite{1hsy-jpcs12,1essay} 
and the emergent spacetime precludes the formation of pocket universes appearing
in the eternal (or chaotic) inflation. 
We also discuss a speculative mechanism to end the inflation by some nonlinear damping through interactions 
between the inflating background and ubiquitous local fluctuations. 
Finally we discuss possible ways to understand our real world $\mathbb{R}^{3,1}$
that is unfortunately beyond our current approach because 
$\mathbb{R}^{3,1}$ does not belong to the family of
(almost) symplectic manifolds.

In Appendix A, we briefly review the mathematical foundation of locally conformal symplectic and 
cosymplectic manifolds that 
correspond to a natural phase space describing the cosmic inflation of our universe. 
In Appendix B, we give a brief exposition of harmonic oscillator with time-dependent mass to illustrate 
how a nonconservative dynamical system with friction can be formulated
by a time-dependent Hamiltonian system which may be useful to understand 
the cosmic inflation as a dynamical system.
In Appendix C, we propose a background-independent formulation of string theory 
in terms of matrix string theory \cite{mst-mst,dvv}. 
We argue that the pseudoholomorphic curve \cite{donaldson} can be generalized to the Hitchin equations 
describing a Higgs bundle \cite{hitchin} by the matrix string theory.

\section{Emergent spacetime from matrix model}

Let us start with a zero-dimensional matrix model with a bunch of $N \times N$ Hermitian matrices,
$\{\phi_a \in \mathcal{A}_N| a= 1, \cdots, 2n\}$, whose action is given by \cite{ikkt}
\begin{equation}\label{ikkt}
    S = - \frac{1}{4} \sum_{a, b=1}^{2n} \Tr [\phi_a, \phi_b]^2.
\end{equation}
We require that the matrix algebra $\mathcal{A}_N$ is associative, from which we get the Jacobi
identity
\begin{equation}\label{jacobi}
    [\phi_a, [\phi_b, \phi_c]] + [\phi_b, [\phi_c, \phi_a]] + [\phi_c, [\phi_a, \phi_b]] = 0.
\end{equation}
We also assume the action principle, from which we yield the equations of motion:
\begin{equation}\label{eom}
   \sum_{b=1}^{2n} [\phi_b, [\phi_a, \phi_b]] = 0.
\end{equation}
We emphasize that we have not introduced any spacetime structure to define the action \eq{ikkt}.
It is enough to suppose the matrix algebra $\mathcal{A}_N$ consisted of a bunch of matrices
which are subject to a few relationships given by Eqs. \eq{jacobi} and \eq{eom}.

First suppose that the vacuum configuration of $\mathcal{A}_N$ is given by
\begin{equation}\label{vaccum}
    \langle \phi_a \rangle_{\mathrm{vac}} = p_a \in \mathcal{A}_N,
\end{equation}
which must be a solution of Eqs. \eq{jacobi} and \eq{eom}. 
In particular, we are interested in the matrix algebra $\mathcal{A}_N$ in the limit $N \to \infty$.
An obvious solution in the limit $N \to \infty$ is given by the Moyal-Heisenberg algebra\footnote{\label{coulomb}The
conventional choice of vacuum in Coulomb branch is given by $[\phi_a, \phi_b]|_{\mathrm{vac}} = 0$
and so $\langle \phi_a \rangle_{\mathrm{vac}} = \mathrm{diag}\big( (\lambda_a)_1, (\lambda_a)_2,
\cdots, (\lambda_a)_N \big)$. However, it turns out (see Section III.C in \cite{1hsy-review}) that,
in order to describe a classical geometry from a background-independent theory,
it is necessary to have a nontrivial vacuum defined by a ``coherent" condensation obeying
the algebra (\ref{moyal-vac}). For this reason, we will choose the Moyal-Heisenberg vacuum
instead of the conventional vacuum. A similar reasoning was also advocated in footnote 2
in Ref. \cite{1hea}.}
\begin{equation}\label{moyal-vac}
    [p_a, p_b] = - i B_{ab},
\end{equation}
where $(B)_{ab} = - l_s^{-2} (\mathbf{1}_n \otimes i \sigma^2)$ is a $2n \times 2n$ constant
symplectic matrix and $l_s$ is a typical length scale set by the vacuum.
A general solution will be generated by considering all possible deformations of the Moyal-Heisenberg
algebra \eq{moyal-vac}. It is assumed to take the form
\begin{equation}\label{gen-sol}
    \phi_a = p_a + \widehat{A}_a \in \mathcal{A}_N,
\end{equation}
obeying the deformed algebra given by
\begin{equation}\label{def-moyal}
    [\phi_a, \phi_b] = - i (B_{ab} - \widehat{F}_{ab}),
\end{equation}
where
\begin{equation}\label{field-st}
    \widehat{F}_{ab} = \partial_a \widehat{A}_b - \partial_b \widehat{A}_a
    - i [\widehat{A}_a, \widehat{A}_b] \in \mathcal{A}_N
\end{equation}
with the definition $\partial_a \equiv \mathrm{ad}_{p_a} = -i [p_a, \cdot]$. For the general matrix
$\phi_a \in \mathcal{A}_N$ to be a solution of Eqs. \eq{jacobi} and \eq{eom}, the set of matrices
$\widehat{F}_{ab} \in \mathcal{A}_N$, called the field strengths of NC $U(1)$ gauge fields $\widehat{A}_a \in \mathcal{A}_N$, must obey the following equations
\begin{eqnarray} \label{bianchi}
&& \widehat{D}_a \widehat{F}_{bc} +  \widehat{D}_b \widehat{F}_{ca} + \widehat{D}_c \widehat{F}_{ab} = 0, \\
\label{nceom}
&&   \sum_{b=1}^{2n} \widehat{D}_b \widehat{F}_{ab} = 0,
\end{eqnarray}
where
\begin{equation}\label{d-fieldst}
    \widehat{D}_a \widehat{F}_{bc} \equiv \mathrm{ad}_{\phi_a} \widehat{F}_{bc}
    = -i [\phi_a, \widehat{F}_{bc}] = - [\phi_a, [\phi_{b}, \phi_c]].
\end{equation}

The algebra $\mathcal{A}_N$ admits a large amount of inner automorphism denoted by $\mathrm{Inn} (\mathcal{A}_N)$.
Note that any automorphism of the matrix algebra $\mathcal{A}_N$ is inner.
Suppose that $\mathcal{A}'_{\widetilde{N}} = \{\phi'_a| a= 1, \cdots, m \}$ is an another matrix algebra
composed of $m$ elements of $\widetilde{N} \times \widetilde{N}$ Hermitian matrices.
We will identify two matrix algebras, i.e. $\mathcal{A}_N \cong \mathcal{A}'_{\widetilde{N}}$
if $m = 2n$ and $\widetilde{N} = N$ and there exists a unitary matrix
$U \in \mathrm{Inn} (\mathcal{A}_N)$ such that $\phi'_a = U \phi_a U^{-1}, \; \forall a = 1, \cdots, 2n$.
It is important to note that the NC algebra $\mathcal{A}_N$ generated by the vacuum operators $p_a$
admits an infinite-dimensional separable Hilbert space
\begin{equation}\label{hilbert}
    \mathcal{H} = \{ | n \rangle | n = 1, \cdots, N \to \infty \},
\end{equation}
that is the Fock space of the Moyal-Heisenberg algebra \eq{moyal-vac}.
As is well-known from quantum mechanics \cite{dirac}, there is a one-to-one correspondence
between the operators in $\mathrm{Hom}(V)$ and the set of $N \times N$ matrices over $\mathbb{C}$
where $V$ is an $N$-dimensional complex vector space. In our case, $V = \mathcal{H}$ is
a Hilbert space and $N = \mathrm{dim} (\mathcal{H}) \to \infty$.
Thus the matrix algebra $\mathcal{A}_N$ can be realized as a Hilbert space representation of
the NC $\star$-algebra
\begin{equation}\label{star-alg}
    \mathcal{A}_\theta = \{\widehat{\phi}_a (y) \in \mathrm{Hom}(\mathcal{H})| a = 1, \cdots, 2n \},
\end{equation}
which is generated by the set of coordinate generators obeying the commutation relation
\begin{equation}\label{1extra-nc2n}
    [y^a, y^b]_\star = i \theta^{ab}. 
\end{equation}
The $\star$-algebra \eq{1extra-nc2n} is related to the Moyal-Heisenberg algebra \eq{def-moyal}
where $p_a = B_{ab} y^b$ and $(\theta)^{ab} = (B^{-1})^{ab} = l_s^2 (\mathbf{1}_n \otimes i \sigma^2)$ 
is a $2n \times 2n$ constant symplectic matrix. 
Let us denote the NC $\star$-algebra  $\mathcal{A}_\theta$ generated by \eq{1extra-nc2n} as $\mathbb{R}^{2n}_{\theta}$.

Given a Hermitian operator $\widehat{\phi}_a (y) \in \mathcal{A}_\theta$,
we have a matrix representation in $\mathcal{H}$ as follows:
\begin{equation}\label{mat-phi}
    \widehat{\phi}_a (y) = \sum_{n,m=1}^\infty |n\rangle \langle n |\widehat{\phi}_a (y) |m \rangle \langle m|
    = \sum_{n,m=1}^\infty (\phi_a)_{nm} |n \rangle \langle m|
\end{equation}
using the completeness of $\mathcal{H}$, i.e. $\sum_{n=1}^\infty | n \rangle \langle n |
= \mathbf{1}_{\mathcal{H}}$. The unitary representation of the operator algebra $\mathcal{A}_\theta$
can thus be understood as a linear transformation acting on an $N$-dimensional
Hilbert space $\mathcal{H}_{N}$:
\begin{equation}\label{unitary-h}
    \mathcal{A}_\theta: \mathcal{H}_{N} \to \mathcal{H}_{N}.
\end{equation}
That is, we have the identification \cite{1japan-matrix,1hsy-epjc09}
\begin{equation}\label{iso-matop}
\mathcal{A}_N \cong \mathrm{End}(\mathcal{H}_{N}) \cong \mathcal{A}_\theta.
\end{equation}

As a result, the inner automorphism $\mathrm{Inn} (\mathcal{A}_N)$ of the matrix algebra $\mathcal{A}_N$
is translated into that of the NC $\star$-algebra $\mathcal{A}_\theta$,
denoted by $\mathrm{Inn} (\mathcal{A}_\theta)$. Its infinitesimal generators consist of an inner
derivation $\mathfrak{D}$ defined by the map \cite{1hsy-jhep09,1hsy-review,1q-emg,1h-stein}
\begin{equation}\label{inner-d}
    \mathcal{A}_\theta \to \mathfrak{D}: \mathcal{O} \mapsto \mathrm{ad}_{\mathcal{O}}
    = - i [\mathcal{O}, \cdot]_\star
\end{equation}
for any operator $\mathcal{O} \in \mathcal{A}_\theta$. Using the Jacobi identity of
the NC $\star$-algebra $\mathcal{A}_\theta$, one can easily verify the Lie algebra homomorphism:
\begin{equation}\label{lie-homo}
    [\mathrm{ad}_{\mathcal{O}_1}, \mathrm{ad}_{\mathcal{O}_2}]
    = - i \mathrm{ad}_{[\mathcal{O}_1, \mathcal{O}_2]_\star}
\end{equation}
for any $\mathcal{O}_1, \mathcal{O}_2 \in \mathcal{A}_\theta$.
In particular, we are interested in the set of
derivations determined by NC gauge fields in Eq. \eq{star-alg}:
\begin{equation}\label{nc-vector}
\{ \widehat{V}_a \equiv \mathrm{ad}_{\widehat{\phi}_a} \in \mathfrak{D}| \widehat{\phi}_a (y)
= p_a + \widehat{A}_a (y) \in \mathcal{A}_\theta, \; a = 1, \cdots, 2n \}.
\end{equation}
In a large-distance limit, i.e. $|\theta| \to 0$, one can expand the NC vector fields $\widehat{V}_a$
using the explicit form of the Moyal $\star$-product. The result takes the form
\begin{equation}\label{polyvector}
  \widehat{V}_a = V^\mu_a (y) \frac{\partial}{\partial y^\mu} + \sum_{p=2}^\infty
  V^{\mu_1 \cdots \mu_p}_a (y) \frac{\partial}{\partial y^{\mu_1}} \cdots
  \frac{\partial}{\partial y^{\mu_p}} \in \mathfrak{D}.
\end{equation}
Thus the NC vector fields in $\mathfrak{D}$ generate an infinite tower of
the so-called polyvector fields \cite{1q-emg}.
Note that the leading term gives rise to the ordinary vector
fields that will be identified with a frame basis associated to the tangent bundle $T\mathcal{M}$
of an emergent manifold $\mathcal{M}$.
If the leading term in \eq{polyvector} already generated the gravitational fields of spin 2,
the higher-order terms would correspond to higher-spin fields with spin $\geq 3$.

Since we have started with a large $N$ matrix model, it is natural to expect that
the IKKT-type matrix model \eq{ikkt} is dual to a higher-dimensional gravity or string theory
according to the large $N$ duality or gauge/gravity duality \cite{1wtaylor}.
The emergent gravity is realized via the gauge-gravity duality as follows \cite{1q-emg}:
\begin{equation}\label{duality-chain}
  \mathcal{A}_N  \quad \Longrightarrow \quad \mathcal{A}_\theta \quad
  \Longrightarrow \quad \mathfrak{D}.
\end{equation}
The gauge theory side of the duality is described by the set of large $N$ matrices
that consists of an associative, but NC, algebra $\mathcal{A}_N$.
By choosing a proper vacuum such as Eq. \eq{vaccum}, a matrix in $\mathcal{A}_N$ is regarded as
a linear representation of an operator acting on a separable Hilbert space $\mathcal{H}$.
That is, the matrix algebra $\mathcal{A}_N$ is realized as a linear representation of
an operator algebra $\mathcal{A}_\theta$ on the Hilbert space $\mathcal{H}$, i.e.,
$\mathcal{A}_N \cong \mathrm{End} (\mathcal{H})$. Consequently the algebra $\mathcal{A}_N$
is isomorphically mapped to the NC $\star$-algebra $\mathcal{A}_\theta$,
as Eq. \eq{mat-phi} has clearly illustrated.
The gravity side of the duality is defined by associating the derivation $\mathfrak{D}$ of
the algebra $\mathcal{A}_\theta$ with a quantized frame bundle $\widehat{\mathfrak{X}}(\mathcal{M})$
of an emergent spacetime manifold $\mathcal{M}$. The noncommutativity of an underlying algebra is
thus crucial to realize the emergent gravity. 
This is the reason why we need the Moyal-Heisenberg vacuum \eq{moyal-vac} instead of
the conventional Coulomb branch vacuum \cite{ly-jkps}. 
After all, in order to describe a quantum geometry properly,
it is necessary to distinguish two types of vacuum in the Coulomb branch: 
diagonalizable vs. non-diagonalizable vacua.

At this stage it is important to understand how (local) coordinates which have been used to 
define the vector fields in $\mathfrak{D}$ arise from matrices in $\mathcal{A}_N$. 
The crux is the isomorphism \eq{iso-matop} between the matrix algebra $\mathcal{A}_N$ and 
the NC $\star$-algebra $\mathcal{A}_\theta$ in the limit $N \to \infty$. 
Here the quantity $|\theta|$ in \eq{1extra-nc2n} plays a role similar to $\hbar$ in quantum mechanics.
Therefore, we will get a classical algebra $C^\infty ({\cal M})$ generated by smooth functions 
on ${\cal M}$ from the NC $\star$-algebra $\mathcal{A}_\theta$ 
when we take a commutative limit, $|\theta| \to 0$. 
Then, given an open set $U \subset {\cal M}$, one can use some local 
functions $(y^1, \cdots, y^{2n}): U \to \mathbb{R}^{2n}$ to define a coordinate chart around $p \in U$. 
Since the underlying functions are smooth, one can introduce infinitesimal quantities such as 
tangent vectors $\frac{\partial}{\partial y^\mu}|_p$ and covectors $dy^\mu |_p$ at $p \in U$ 
associated with the given coordinate system. Note that, if we had chosen a diagonalized vacuum (see footnote \ref{coulomb}) 
instead of the non-diagonalizable vacuum \eq{moyal-vac}, 
the existence of such continuous variables and infinitesimal values would not be guaranteed even in the limit $N \to \infty$.

Recognizing the intrinsic locality is crucial when grasping the emergence of geometry 
through the duality chain in Eq. \eq{duality-chain}. It is necessary
to consider patching or gluing together the local constructions to form
a set of global quantities. For this purpose, the concept of sheaf may be essential
because it makes it possible to reconstruct global data starting from open sets of
locally defined data \cite{gh-book}. We provide a succinct overview of this feature, 
as it has already been comprehensively discussed in Ref. \cite{1q-emg}. 
Its characteristic feature becomes transparent
when the commutative limit, i.e. $|\theta| \to 0$, is taken into account.
In this limit, the NC $\star$-algebra $\mathcal{A}_\theta$ reduces to a Poisson algebra
$\mathfrak{P}^{(i)} = (C^\infty(U_i), \{-,-\}_\theta)$ defined on a local patch $U_i \subset M$ in an
open covering $M = \bigcup_{i \in I} U_i$.\footnote{We will often use the symbol $M$ to denote a generic manifold 
whereas the symbol ${\cal M}$ has been used to emphasize an emergent manifold.} 
The Poisson algebra $\mathfrak{P}^{(i)}$ arises as follows.
Let $L \to M$ be a line bundle over $M$ whose connection is denoted by $\mathcal{A}$.
We assume that the curvature $\mathcal{F}$ of
the line bundle $L$ is a {\it nondegenerate}, closed two-form. Therefore we identify
the curvature two-form $\mathcal{F} = d \mathcal{A}$ with a symplectic structure of $M$.
On an open neighborhood $U_i \subset M$, it is possible to represent $\mathcal{F}^{(i)} = B + F^{(i)}$
where $F^{(i)} = d A^{(i)}$ and $B$ is the constant symplectic two-form already introduced
in Eq. \eq{moyal-vac}. Consider a chart $(U_i, \phi_{(i)})$ where $\phi_{(i)} \in \mathrm{Diff}(U_i)$
is a local trivialization of the line bundle $L$ over the open subset $U_i$
obeying $\phi_{(i)}^* (\mathcal{F}^{(i)}) = B$. 
A local chart is guaranteed to exist thanks to either the Darboux theorem or the Moser lemma
in symplectic geometry \cite{sg-book} and the local coordinate chart
obeying $\phi_{(i)}^* (\mathcal{F}^{(i)}) = B$ is called Darboux coordinates.
Thus the line bundle $L \to M$ corresponds to a dynamical symplectic manifold $(M, \mathcal{F})$
where $\mathcal{F} = B + dA$. The dynamical system is locally described by the Poisson
algebra $\mathfrak{P}^{(i)} = (C^\infty(U_i), \{-,-\}_\theta)$ in which the vector space $C^\infty(U_i)$
is formed by the set of Darboux transformations $\phi_{(i)} \in \mathrm{Diff}(U_i)$ equipped
with the Poisson bracket defined by the Poisson bivector $\theta = B^{-1} \in \Gamma(\Lambda^2 TM)$.

Consider a collection of local charts to make an atlas $\{(U_i, \phi_{(i)})\}$ on $M = \bigcup_{i \in I} U_i$
and complete the atlas by gluing these charts on their overlaps.
To be precise, suppose that $(U_i, \phi_{(i)})$ and $(U_j, \phi_{(j)})$ are two coordinate charts
and $F^{(i)} = d A^{(i)}$ and $F^{(j)} = d A^{(j)}$ are local curvature two-forms
on $U_i$ and $U_j$, respectively. We choose the coordinate maps $\phi_{(i)} \in \mathrm{Diff}(U_i)$
and $\phi_{(j)} \in \mathrm{Diff}(U_j)$ such that $\phi_{(i)}^* (B + F^{(i)}) = B$ and
$\phi_{(j)}^* (B + F^{(j)}) = B$. On an intersection $U_i \cap U_j$, the local data
$(A^{(i)}, \phi_{(i)})$ and $(A^{(j)}, \phi_{(j)})$ on Darboux charts $(U_i, \phi_{(i)})$
and $(U_j, \phi_{(j)})$, respectively, are glued together by \cite{nc-glue}
\begin{eqnarray}\label{glue-g}
&& A^{(j)} = A^{(i)} + d \lambda^{(ji)}, \\
\label{glue-diff}
&& \phi_{(ji)} = \phi_{(j)} \circ \phi^{-1}_{(i)},
\end{eqnarray}
where $\phi_{(ji)} \in \mathrm{Diff}(U_i \cap U_j)$ is a symplectomorphism on $U_i \cap U_j$ generated
by a Himiltonian vector field $X_{\lambda^{(ji)}}$ satisfying $\iota (X_{\lambda^{(ji)}}) B
+ d \lambda^{(ji)} = 0$. We sometimes denote the interior product $\iota_X$ by $\iota(X)$
for a notational convenience. Similarly, we can glue the local Poisson algebras $\mathfrak{P}^{(i)}$
to form a globally defined Poisson algebra $\mathfrak{P} = \bigcup_{i \in I} \mathfrak{P}^{(i)}$.
The global vector fields $V_a = V^\mu_a (y) \frac{\partial}{\partial y^\mu} \in \Gamma(T\mathcal{M}),
\; a = 1, \cdots, 2n$, in Eq. \eq{polyvector} can be obtained by applying a similar globalization
to the derivation $\mathfrak{D}$, which form a linearly independent basis of the tangent bundle
$T \mathcal{M}$ of a $2n$-dimensional emergent manifold $\mathcal{M}$. As a consequence,
the set of global vector fields $\mathfrak{X}(\mathcal{M}) = \{ V_a| a = 1, \cdots, 2n \}$
results from the globally defined Poisson algebra $\mathfrak{P}$ \cite{1q-emg}.

The vector fields $V_a \in \mathfrak{X}(\mathcal{M})$ are related to an orthonormal frame,
the so-called vielbeins $E_a \in \Gamma(T\mathcal{M})$, in general relativity by the relation
\begin{equation}\label{ve-rel}
    V_a = \lambda E_a, \qquad  a = 1, \cdots, 2n.
\end{equation}
The conformal factor $\lambda \in C^\infty (\mathcal{M})$ is determined by imposing the condition that
the vector fields $V_a$ preserve a volume form
\begin{equation}\label{vol-form}
    \nu = \lambda^2 v^1 \wedge \cdots \wedge v^{2n},
\end{equation}
where $v^a = v^a_\mu (y) dy^\mu \in \Gamma(T^* \mathcal{M})$ are coframes dual to $V_a$, i.e.,
$\langle v^a, V_b \rangle = \delta^a_b$. This means that the vector fields $V_a$ obey the conditions
\begin{equation}\label{vol-cond}
    \mathcal{L}_{V_a} \nu = \big( \nabla \cdot V_a + (2-2n) V_a \ln \lambda \big) \nu = 0,
    \qquad \forall a = 1, \cdots, 2n,
\end{equation}
where $\mathcal{L}_X  = \iota_X d + d \iota_X$ is the Lie derivative with respect to a vector field $X$.
Note that a symplectic manifold always admits such volume-preserving vector fields.
(See Appendix B in \cite{1q-emg}.)
Together with the volume-preserving condition \eq{vol-cond}, the relation \eq{ve-rel}
completely determines a $2n$-dimensional Riemannian manifold $\mathcal{M}$
whose metric is given by \cite{1hsy-jhep09,1hsy-review,1q-emg}
\begin{eqnarray} \label{em-metric}
  ds^2 &=& \mathcal{G}_{\mu\nu} (x) dx^\mu \otimes dx^\nu = e^a \otimes e^a \nonumber \\
  &=& \lambda^2 v^a \otimes v^a = \lambda^2 v^a_\mu (y) v^a_\nu (y) dy^\mu \otimes dy^\nu,
\end{eqnarray}
where $e^a = e^a_\mu (x) dx^\mu = \lambda v^a \in \Gamma(T^* \mathcal{M})$ are orthonormal one-forms
on $\mathcal{M}$. After all, the $2n$-dimensional Riemannian manifold $\mathcal{M}$ is emergent
from the commutative limit of polyvector fields $\widehat{V}_a = V_a + \mathcal{O} (\theta^2)
\in \mathfrak{D}$ derived from NC $U(1)$ gauge fields.

So far we have discussed the emergence of spaces only. However, the theory of relativity dictates
that space and time must be coalesced into the form of Minkowski spacetime in a locally inertial
frame. Hence, if general relativity is realized from a NC $\star$-algebra $\mathcal{A}_\theta$,
it is necessary to put space and time on an equal footing in the NC $\star$-algebra $\mathcal{A}_\theta$.
If space is emergent, so should time.
Thus, an important problem is how to realize the emergence of ``time."
However, any physical theory that we know does not treat time as a dynamical variable. 
Therefore, we assert that the concept of emergent time needs to be understood differently from emergent spaces.
(We will later discuss a perplexing problem that arises when we promote time to a ``dynamical" variable.) 
Quantum mechanics imparts a valuable insight, emphasizing the intricate relationship between 
the definition of (particle) time and the dynamics inherent in the system. 
In quantum mechanics, the time evolution of a dynamical system is defined
as an inner automorphism of NC algebra $\mathcal{A}_\hbar$ generated by the NC phase space
\begin{equation}\label{nc-phase}
    [x^i, x^j] = 0, \qquad  [x^i, p_j] = i \hbar \delta^i_j, \qquad i, j = 1, \cdots, n.
\end{equation}
The time evolution for an observable $f \in \mathcal{A}_\hbar$ is simply an inner derivation
of $\mathcal{A}_\hbar$ given by
\begin{equation}\label{h-eq}
 \frac{df}{dt} = \frac{i}{\hbar} [H, f],
\end{equation}
where $H$ is a Hamiltonian operator of the dynamical system and will be identified 
with a temporal gauge field $A_0$, i.e. $H = - A_0$, in matrix quantum mechanics.
The integral of Eq. \eq{h-eq} is simply a unitary transformation of 
the observable $f \in \mathcal{A}_\hbar$:
\begin{equation}\label{time-unit}
 f(t) = U(t) f(0) U(t)^\dagger,
\end{equation}
where $U(t) = e^{\frac{iH t}{\hbar}}$ is a unitary operator. 
For a quantum dynamical system that has a classical analogue, Eq. \eq{time-unit} implies that 
unitary transformations in the quantum theory are analogue of canonical (or contact) transformations 
in the classical theory (see \S 26 Unitary transformations in \cite{dirac}).

Given a symplectic form $\omega = \sum_{i=1}^n dx^i \wedge dp_i$ on phase space, one can introduce 
a Hamiltonian vector field $X_H$ defined by $\iota_{X_H} \omega = dH$. 
The one-parameter family of canonical transformations can then be thought of as ``Hamiltonian flow" 
on phase space:
\begin{equation} \label{ham-flow}
\big( X^i (x,p; t) = x^i + t X_H (x^i), P_i (x,p; t) = p_i + t X_H (p_i) \big).    
\end{equation}
According to this active viewpoint, the canonical transformation takes one point in the phase space,
$(x^i, p_i)$, to another point in the same phase space, $\big( X^i (x,p; t), P_i (x,p; t) \big)$. 
Correspondingly, the point at time $t$ can be understood 
as a one-parameter family of deformations (or changes) generated by a smooth function $H=H(x,p)$. 
We will define the concept of emergent time based on this perspective.

A remarkable picture, as observed by Feynman \cite{feyn-dy}, Souriau, and Sternberg \cite{sstern},
is that the physical
forces such as the electromagnetic, weak and strong forces, can be realized as the deformations
of an underlying vacuum algebra such as Eq. \eq{nc-phase}. For example, the most general deformation
of the Heisenberg algebra \eq{nc-phase} within the {\it associative} algebra $\mathcal{A}_\hbar$
is given by
\begin{equation}\label{feynman}
    x^i \to x^i, \qquad p_i \to p_i + A_i (x, t), \qquad H \to H + A_0 (x, t),
\end{equation}
where $(A_0, A_i) (x, t)$ must be electromagnetic gauge fields. Then the time evolution of
a particle system under a time-dependent external force is given by
\begin{equation}\label{th-eq}
\frac{df}{dt} = \frac{\partial f}{\partial t} + \frac{i}{\hbar} [H, f].
\end{equation}

Note that the construction of the NC algebra $\mathcal{A}_N$ or $\mathcal{A}_\theta$ bears
a close parallel to quantum mechanics. The former is based on the NC space \eq{1extra-nc2n}
while the latter is based on the NC phase space \eq{nc-phase}.
The NC $U(1)$ gauge fields in Eq. \eq{gen-sol} act as deformations
of the vacuum algebra \eq{moyal-vac} in the matrix algebra $\mathcal{A}_N$,
similarly to Eq. \eq{feynman} in the quantum algebra $\mathcal{A}_\hbar$.
Therefore we can apply the same philosophy to the NC algebra $\mathcal{A}_N$ or $\mathcal{A}_\theta$
to define a dynamical system based on the Moyal-Heisenberg algebra \eq{moyal-vac}.
In other words, we can consider a one-parameter family of deformations of zero-dimensional matrices
which is parameterized by the coordinate $t$. Then the one-parameter family of deformations
characterized by \eq{gen-sol} can be regarded as
the time evolution of a dynamical system. For this purpose, we extend the NC algebra
$\mathcal{A}_\theta$ to $\mathcal{A}^1_\theta \equiv \mathcal{A}_\theta \big( C^\infty (\mathbb{R}) \big)
= C^\infty (\mathbb{R}) \otimes \mathcal{A}_\theta$ whose generic element
takes the form
\begin{equation}\label{contact-f}
    \widehat{f}(t, y) \in \mathcal{A}^1_\theta.
\end{equation}
The matrix representation \eq{mat-phi} is then replaced by
\begin{equation}\label{mat-time}
    \widehat{f}(t, y) = \sum_{n,m=1}^\infty |n\rangle \langle n |\widehat{f}(t, y) |m \rangle \langle m|
    = \sum_{n,m=1}^\infty f_{nm} (t) |n \rangle \langle m|
\end{equation}
where $f_{nm} (t):= [f(t)]_{nm}$ are elements of a matrix $f(t)$ in $\mathcal{A}^1_N \equiv
\mathcal{A}_N \big( C^\infty (\mathbb{R}) \big) = C^\infty (\mathbb{R}) \otimes \mathcal{A}_N$
as a representation of the observable \eq{contact-f} on the Hilbert space \eq{hilbert}.
As the Heisenberg equation \eq{th-eq} in quantum mechanics suggests, the evolution equation
for an observable $\widehat{f}(t, y) \in \mathcal{A}^1_\theta$ in the Heisenberg picture is defined by
\begin{equation}\label{h-equation}
 \frac{d\widehat{f}(t, y)}{dt} = \frac{\partial \widehat{f}(t, y)}{\partial t}
 -i [\widehat{A}_0 (t, y), \widehat{f}(t, y)]_\star \equiv \widehat{D}_0 \widehat{f}(t, y)
\end{equation}
where we denoted the local Hamiltonian density by
\begin{equation}\label{local-ham}
   \widehat{H} (t, y) \equiv - \widehat{A}_0 (t, y)  \in \mathcal{A}^1_\theta.
\end{equation}
Note that
\begin{equation}\label{cov-2n}
    -i [\phi_a, \widehat{f}(t)] = \partial_a \widehat{f}(t, y)
    - i [\widehat{A}_a (t, y), \widehat{f}(t, y)]_\star \equiv \widehat{D}_a \widehat{f}(t, y),
\end{equation}
where the representation \eq{mat-time} has been employed.
Then one can see that the inner automorphism $\mathrm{Inn} (\mathcal{A}_\theta)$ of $\mathcal{A}_\theta$
can be lifted to the automorphism of $\mathcal{A}^1_\theta$ given by
\begin{eqnarray} \label{time-auto}
&& \widehat{A}_0 (t, y) \to \widehat{U} (t,y) \star \frac{\partial \widehat{U}^{-1} (t, y)}{\partial t} +
\widehat{U} (t,y) \star \widehat{A}_0 (t, y) \star \widehat{U}^{-1} (t, y), \\
\label{space-auto}
&& \widehat{A}_a (t, y) \to \widehat{U} (t,y) \star \frac{\partial \widehat{U}^{-1} (t, y)}{\partial y^a} + \widehat{U} (t,y) \star \widehat{A}_a (t, y) \star \widehat{U}^{-1} (t, y),
\end{eqnarray}
where $\widehat{U} (t,y) = e^{i \widehat{\lambda} (t, y)}_\star$ with $\widehat{\lambda} (t, y)
\in \mathcal{A}^1_\theta$. It is obvious that the above automorphism is nothing but
the gauge transformation for NC $U(1)$ gauge fields in $(2n+1)$-dimensions \cite{ncft-sw}.

Our leitmotif is that a consistent theory of quantum gravity should be background-independent,
so that it should not presuppose any spacetime background on which fundamental processes develop.
Hence the background-independent theory must provide a mechanism of spacetime generation such that
every spacetime structure including the flat spacetime arises as a solution of the theory itself.
A zero-dimensional matrix model such as Eq. \eq{ikkt} is the most natural candidate for such a background-independent theory 
because it does not have to assume the prior existence of spacetime to define the theory.

Then, how can Minkowski spacetime also emerge as a solution of an underlying background-independent theory?
We emphasize again that the NC nature of the vacuum solution, e.g. Eq. \eq{moyal-vac},
is essential to realize the large $N$ duality via the duality chain \eq{duality-chain}.
A profound feature is that the background-independent theory is intrinsically dynamical
because the space of all possible solutions is generated by generic deformations of a primitive vacuum such as Eq. \eq{moyal-vac} \cite{1q-emg}.
We contend that the dynamics governed by the Moyal-Heisenberg vacuum \eq{vaccum} is characterized 
by the NC algebra $\mathcal{A}^1_N = \mathcal{A}_N \big( C^\infty (\mathbb{R}) \big)
= C^\infty (\mathbb{R}) \otimes \mathcal{A}_N$. One may regard $\mathcal{A}^1_N$ as a one-parameter
family of deformations of the algebra $\mathcal{A}_N$.
In this case we can generalize the duality chain \eq{duality-chain} to realize the ``time-dependent"
gauge/gravity duality as follows:
\begin{equation}\label{tduality-chain}
  \mathcal{A}^1_N  \quad \Longrightarrow \quad \mathcal{A}^1_\theta \quad
  \Longrightarrow \quad \mathfrak{D}^1.
\end{equation}
It is well-known \cite{azam} that in the case of $\mathcal{A}^1_N$ or $\mathcal{A}^1_\theta$,
the module of its derivations can be written as a direct sum of the submodules of horizontal
and inner derivations:
\begin{equation}\label{1deriv}
  \mathfrak{D}^1 = \mathrm{Hor}(\mathcal{A}^1_N) \oplus \mathfrak{D} (\mathcal{A}^1_N) \cong
  \mathrm{Hor}(\mathcal{A}^1_\theta) \oplus \mathfrak{D} (\mathcal{A}^1_\theta)
\end{equation}
where horizontal derivation is a lifting of smooth vector fields on $\mathbb{R}$ onto $\mathcal{A}^1_N$
or $\mathcal{A}^1_\theta$ and is locally generated by a vector field
\begin{equation}\label{time-vec}
    g(t, y) \frac{\partial}{\partial t} \in \mathrm{Hor}(\mathcal{A}^1_\theta).
\end{equation}
The inner derivation $\mathfrak{D} (\mathcal{A}^1_\theta)$ is defined by lifting the NC vector fields
in Eq. \eq{nc-vector} onto $\mathcal{A}^1_\theta$ and generated by
\begin{equation}\label{nc-tvector}
\{ \widehat{V}_a (t) \equiv \mathrm{ad}_{\widehat{\phi}_a} \in \mathfrak{D} (\mathcal{A}^1_\theta)|
\widehat{\phi}_a (t, y) = p_a + \widehat{A}_a (t, y) \in \mathcal{A}^1_\theta, \; a = 1, \cdots, 2n \}
\end{equation}
and
\begin{equation}\label{time-vector}
\Big\{ \widehat{V}_0 (t) - \frac{\partial}{\partial t} \equiv \mathrm{ad}_{\widehat{A}_0} \in \mathfrak{D} (\mathcal{A}^1_\theta)| \widehat{A}_0 (t, y) \in \mathcal{A}^1_\theta \Big\}.
\end{equation}
It might be remarked that the definition of the time-like vector field $\widehat{V}_0 (t)$ is
motivated by the quantum Hamilton's equation \eq{h-equation}, i.e.,
\begin{equation}\label{timelike-vec}
  \widehat{V}_0 (t) := \frac{d}{dt}.
\end{equation}
Consequently, the module of the derivations of the NC algebra $\mathcal{A}^1_\theta$ is given by
\begin{equation}\label{time-derivation}
\mathfrak{D}^1 = \Big\{ \widehat{V}_A (t) = \big( \widehat{V}_0, \widehat{V}_a \big) (t) |
\widehat{V}_0 (t) = \frac{\partial}{\partial t} + \mathrm{ad}_{\widehat{A}_0}, \;
\widehat{V}_a (t) = \mathrm{ad}_{\widehat{\phi}_a}, \; A=0, 1, \cdots, 2n \Big\}.
\end{equation}

In the commutative limit, $|\theta| \to 0$, the time-dependent polyvector fields $\widehat{V}_A (t)$
in $\mathfrak{D}^1$ take the following form
\begin{eqnarray}\label{time-polyvec}
&&  \widehat{V}_0 (t) = \frac{\partial}{\partial t} + A_0^\mu (t, y) \frac{\partial}{\partial y^\mu} + \sum_{p=2}^\infty A^{\mu_1 \cdots \mu_p}_0 (t, y) \frac{\partial}{\partial y^{\mu_1}} \cdots
  \frac{\partial}{\partial y^{\mu_p}}, \\
  \label{space-polyvec}
&&  \widehat{V}_a (t) = V_a^\mu (t, y) \frac{\partial}{\partial y^\mu} + \sum_{p=2}^\infty
  V^{\mu_1 \cdots \mu_p}_a (t, y) \frac{\partial}{\partial y^{\mu_1}} \cdots
  \frac{\partial}{\partial y^{\mu_p}}.
\end{eqnarray}
Let us truncate the above polyvector fields to ordinary vector fields given by
\begin{equation}\label{lorentzian-vec}
 \mathfrak{X}(\mathcal{M}) = \Big\{ V_A = V_A^M (t, y) \frac{\partial}{\partial X^M}|
 A, M = 0, 1, \cdots, 2n \Big\}
\end{equation}
where $V_A^0 = \delta^0_A$ and $X^M = (t, y^\mu)$ are local coordinates on an emergent {\it Lorentzian}
manifold $\mathcal{M}$ of $(2n+1)$-dimensions. The orthonormal vielbeins on $T\mathcal{M}$ are then
obtained by the prescription 
\begin{equation}\label{lorentz-iviel}
    (V_0, V_a) = (E_0, \lambda E_a) \in \Gamma(T\mathcal{M}). 
\end{equation}
The dual orthonormal basis on $T^* \mathcal{M}$ is defined by the relation $\langle v^A, V_B \rangle = \delta^A_B$ and it is given by $v^A = (v^0, v^a) = \left(dt, 
v^a_\mu  \big( dy^\mu - A_0^\mu (t, y) \big) \right)$ where $v^a_\mu V^\mu_b = \delta^a_b$.
From Eq. \eq{lorentz-iviel}, we have
\begin{equation}\label{lorentz-viel}
    (e^0, e^a) = ( v^0, \lambda v^a) \in \Gamma(T^* \mathcal{M}).
\end{equation}
The conformal factor $\lambda \in C^\infty (\mathcal{M})$ is similarly determined by
the volume-preserving condition
\begin{equation}\label{tvol-cond}
    \mathcal{L}_{V_A} \nu_t = \big( \nabla \cdot V_A + (2-2n) V_A \ln \lambda \big) \nu_t = 0,
    \qquad \forall A = 0, 1, \cdots, 2n.
\end{equation}
The above condition explicitly reads as
\begin{equation}\label{vol-new}
    \frac{\partial \rho}{\partial t} +  \partial_\mu (\rho A_0^\mu) = 0 \quad \& \quad
    \partial_\mu (\rho V^\mu_a) = 0,
\end{equation}
where $\rho = \lambda^2 \det v^a_\mu$ and
\begin{equation}\label{vol-t}
    \nu_t \equiv dt \wedge \nu = \lambda^2  dt \wedge v^1 \wedge \cdots \wedge v^{2n}
\end{equation}
is a $(2n+1)$-dimensional volume form on $\mathcal{M}$.
If the structure equation of vector fields $V_A \in \Gamma(T\mathcal{M})$ is defined by
\begin{equation}\label{vstr-eq}
    [V_A, V_B] = - {g_{AB}}^C V_C,
\end{equation}
the volume-preserving condition \eq{tvol-cond} can equivalently be written as \cite{1hsy-review,ly-jkps}
\begin{equation}\label{vol-lamb}
    {g_{BA}}^B = V_A \ln \lambda^2.
\end{equation}
In the end, the Lorentzian metric on a $(2n+1)$-dimensional spacetime manifold $\mathcal{M}$
is given by \cite{1hsy-jhep09,1q-emg}
\begin{eqnarray}\label{eml-metric}
    ds^2 &=& \mathcal{G}_{MN} (X) dX^M \otimes dX^N = \eta_{AB} e^A \otimes e^B \nonumber \\
    &=& - v^0 \otimes v^0 + \lambda^2 v^a \otimes v^a = -dt^2 + \lambda^2 v^a_\mu v^a_\nu (dy^\mu - \mathbf{A}^\mu)
    (dy^\nu - \mathbf{A}^\nu)
\end{eqnarray}
where $\mathbf{A}^\mu := A_0^\mu (t, y) dt$.

It should be noted that the time evolution \eq{timelike-vec} for a general time-dependent system
is not completely generated by an inner automorphism since $\mathrm{Hor}(\mathcal{A}^1_\theta)$ is not
an inner but outer derivation. This happens since the time variable $t$ is single.
Thus one may extend the phase space by introducing a conjugate variable $H$ of $t$ so that
the extended phase space becomes a symplectic manifold. Then it is well-known \cite{sg-book} that
the time evolution of a time-dependent system can be defined by the inner automorphism
of the extended phase space whose extended Poisson bivector is given by
\begin{equation}\label{ext-poisson}
    \vartheta = \theta + \frac{\partial}{\partial t} \bigwedge \frac{\partial}{\partial H}
\end{equation}
where
\begin{equation}\label{poisson-theta}
    \theta = \frac{1}{2} \theta^{\mu\nu} \frac{\partial}{\partial y^\mu} \bigwedge
    \frac{\partial}{\partial y^\nu}
\end{equation}
is the original Poisson bivector related to the NC space \eq{1extra-nc2n}.
As a result, one can see \cite{1hsy-review} that the temporal vector field \eq{timelike-vec}
is realized as a generalized Hamiltonian vector field defined by
\begin{equation}\label{inner-time}
    V_0 = \mathcal{X}_H = - \vartheta (dH-dA_0) = \frac{\partial}{\partial t} + X_H
\end{equation}
where $X_H = \theta (dA_0)$ is the original Hamiltonian vector field which is a classical part
of the inner derivation $\mathrm{ad}_{\widehat{A}_0} = X_H + \mathcal{O} (\theta^2) \in \mathfrak{D}(\mathcal{A}^1_\theta)$. 
However, we must bear the cost associated with the extension of the phase space. 
In the extended phase space, the time $t$ is now promoted to a dynamical variable whereas
it was simply an affine parameter describing a Hamiltonian flow in the old phase space.
Then the extended Poisson structure \eq{ext-poisson} raises a serious issue
whether the time variable for a general time-dependent system might also be quantized;
in other words, time also becomes an operator obeying the commutation relation $[t,H] = - i$.
Then it becomes difficult to defend the causality of physical theories.
We want to refrain from addressing this abstruse issue, since it persists as a challenging open problem, 
even within the realm of quantum mechanics.

We address the time issue through a more pragmatic approach.\footnote{But we point out that 
there is another approach for the emergent time where time is regarded as a dynamical variable, 
for example, Ref. \cite{1jun-sang}. 
Therefore our approach for the emergent time must be considered as an alternative viewpoint.}
In mechanical systems, the time is defined through a contact structure \cite{contact-book,duggal}.
Suppose that $(M, B \equiv \theta^{-1})$ is the original symplectic manifold. 
Now we consider a contact manifold $(\mathbb{R} \times M, \widetilde{B})$
where $\widetilde{B} = \pi^*_2 B$ is defined by the projection $\pi_2: \mathbb{R} \times M \to M,
\; \pi_2 (t, x) = x$ \cite{sg-book}.
We define the concept of spacetime in emergent gravity through the contact
manifold $(\mathbb{R} \times M, \widetilde{B})$ in the sense that the derivations
in Eq. \eq{time-derivation} can be obtained by quantizing the contact
manifold $(\mathbb{R} \times M, \widetilde{B})$.
Indeed it is shown in Appendix A that the time-like vector field $V_0$ in Eq. \eq{inner-time} arises
as a Hamiltonian vector field of a cosymplectic manifold whose particular class is a contact manifold.
Note that the emergent geometry described by the metric \eq{eml-metric} respects
the (local) Lorentz symmetry. One can see that
the Lorentzian manifold $\mathcal{M}$ becomes the Minkowski spacetime on a local Darboux chart
in which all fluctuations die out, i.e., $v_\mu^a \to \delta^a_\mu, \; \mathbf{A}^\mu \to 0$, so
$\lambda \to 1$. We have to emphasize that the vacuum algebra
responsible for the emergence of the Minkowski spacetime is the Moyal-Heisenberg algebra \eq{moyal-vac}.
Many surprising results will immediately come out from this dynamical origin of
the flat spacetime \cite{1hsy-jpcs12,1hsy-jhep09,1hsy-review}, which is absent in general relativity.

We close this section by observing that the quantized version of the contact
manifold $(\mathbb{R} \times M, \widetilde{B})$ is described by a matrix quantum mechanics (MQM) whose action is given by
\begin{equation}\label{bfss}
    S = \frac{1}{g_{YM}^2} \int dt \Tr \Big( \frac{1}{2}  (D_0 \phi_a)^2
    + \frac{1}{4}[\phi_a, \phi_b]^2 \Big),
\end{equation}
where $D_0 \phi_a = \frac{\partial \phi_a}{\partial t} - i [A_0, \phi_a]$.
The equations of motion for the matrix action \eq{bfss} are given by
\begin{equation}\label{eom-mqm}
    D_0^2 \phi_a + [\phi_b, [\phi_a, \phi_b]] = 0,
\end{equation}
which must be supplemented with the Gauss constraint
\begin{equation}\label{gauss-mqm}
    [\phi_a, D_0 \phi_a] = 0.
\end{equation}
We interpret the matrix model (\ref{bfss}) as a Hamiltonian system
of the IKKT matrix model whose action is given by Eq. \eq{ikkt}.
Note that the original BFSS matrix model \cite{bfss}
contains 9 adjoint scalar fields while the action \eq{bfss} has even number of adjoint scalar fields.
For the former case, we do not know how to realize the adjoint scalar fields
as a matrix representation of NC $U(1)$ gauge fields on a Hilbert space like as \eq{mat-time}.
Even it may be nontrivial to construct the Hilbert space
because the M-theory is involved with a 3-form instead of symplectic 2-form.
For the latter case, on the other hand, the previous Moyal-Heisenberg vacuum \eq{vaccum}
is naturally extended to the vacuum configuration
of $\mathcal{A}_N^1$ given by
\begin{equation}\label{1-vaccum}
    \langle \phi_a \rangle_{\mathrm{vac}} = p_a, \qquad
    \langle A_0 \rangle_{\mathrm{vac}} = - \mathcal{E},
\end{equation}
where the vacuum moduli $p_a \in \mathcal{A}^1_N$ satisfy the commutation relation \eq{moyal-vac}
and $\mathcal{E}$ is a constant vacuum energy density proportional to the identity matrix.
We consider all possible deformations of the vacuum \eq{1-vaccum} and parameterize them as
\begin{equation}\label{1-deform}
    \widehat{\phi}_A (t, y) = p_A + \widehat{A}_A (t, y) \in \mathcal{A}^1_\theta,
\end{equation}
where $p_0 = i \frac{\partial}{\partial t} - \mathcal{E}$ and 
the isomorphism \eq{mat-time} between $\mathcal{A}^1_N$ and $\mathcal{A}^1_\theta$ was used.
Note that
\begin{equation}\label{1-commrel}
   [\widehat{\phi}_A, \widehat{\phi}_B]_\star = - i \big( B_{AB} - \widehat{F}_{AB} \big),
\end{equation}
where
\begin{equation}\label{1-fieldst}
 \widehat{F}_{AB} = \partial_A \widehat{A}_B - \partial_B \widehat{A}_A
  - i [\widehat{A}_A, \widehat{A}_B]_\star \in \mathcal{A}^1_\theta
\end{equation}
and
$$B_{AB} = \left(
    \begin{array}{cc}
      0 & 0 \\
      0 & B_{ab} \\
    \end{array}
  \right).
$$
Plugging the fluctuations \eq{1-deform} into the action \eq{bfss} leads to a $(2n+1)$-dimensional
NC $U(1)$ gauge theory with the action \cite{1hea,1hsy-epjc09}
\begin{eqnarray}\label{bfss-u1}
    S &=& \frac{1}{g_{YM}^2} \int dt \Tr \Big( \frac{1}{2}  (D_0 \phi_a)^2
    + \frac{1}{4}[\phi_a, \phi_b]^2 \Big) \nonumber \\
    &=& - \frac{1}{4 G_{YM}^2} \int d^{2n+1} y \big( \widehat{F}_{AB} - B_{AB} \big)^2,
\end{eqnarray}
where $G_{YM}^2 = (2\pi)^n |\mathrm{Pf} \theta| g_{YM}^2$ is the $(2n+1)$-dimensional
gauge coupling constant. By applying the duality chain \eq{tduality-chain}
to time-dependent matrices in $\mathcal{A}^1_N$,
it is straightforward to derive the module $\mathfrak{D}^1$
in Eq. \eq{time-derivation} from the large $N$ matrices or NC $U(1)$ gauge fields
in the action \eq{bfss-u1}. A Lorentzian spacetime described by the metric \eq{eml-metric} corresponds
to a classical geometry derived from the NC module $\mathfrak{D}^1$ \cite{1q-emg}.

\section{Cosmic inflation from time-dependent matrices}

From now on we will focus on the matrix quantum mechanics (MQM) 
to address the background-independent formulation of cosmic inflation.
Let us rewrite the action \eq{bfss} as the form
\begin{eqnarray} \label{1mqm}
 S = \frac{1}{4 g^2} \int dt \, \eta^{AC} \eta^{BD} \mathrm{Tr} [\phi_A, \phi_B] [\phi_C, \phi_D],
\end{eqnarray}
where $\phi_0 \equiv i D_0 = i \frac{\partial}{\partial t} + A_0 (t), \; \phi_A (t) = (\phi_0, \phi_a) (t)$
and $\eta^{AB} = \mathrm{diag} (-1, 1, \cdots, 1)$, $A, B = 0, 1, \cdots, 2n.$
With the definition of the symbol $\eta^{AB}$, it is easy to see that the matrix action \eq{1mqm}
has a global automorphism given by
\begin{equation}\label{1poincare-auto}
    \phi_A \to \phi'_A = {\Lambda_A}^B \phi_B + c_A
\end{equation}
if ${\Lambda_A}^B$ is a rotation in $SO(2n,1)$ and $c_A$ are constants proportional
to the identity matrix. It will be shown later that the global symmetry \eq{1poincare-auto}
is responsible for the Poincar\'e symmetry of flat spacetime emergent from a vacuum
in the Coulomb branch of MQM and so will be called the Poincar\'e automorphism.
We remark that the time $t$ in the action \eq{1mqm} is not a dynamical variable but an affine parameter.
The concept of emergent time has been defined in the previous section by considering a one-parameter
family of deformations of zero-dimensional matrices which is parameterized by the coordinate $t$.
The one-parameter family of deformations can then be regarded as the time evolution of
a dynamical system. In this context, the one-dimensional matrix model \eq{1mqm} can be interpreted as
a Hamiltonian system of a zero-dimensional (e.g., IKKT) matrix model \cite{1q-emg}. 
A close analogy with quantum mechanics implies that
the concept of emergent time is derived from the time evolution of the dynamical system.
Although spatial coordinates and time are introduced in different ways, Eq. \eq{1poincare-auto} implies 
that they are connected by Lorentz transformations and coalesced into the form of Minkowski spacetime 
in a locally inertial frame.

The duality chain \eq{tduality-chain} implies that the gravitational variables such as
vielbeins in general relativity arise from the commutative limit of NC $U(1)$ gauge fields. 
Then one may ask where the Minkowski spacetime comes from.
Let us look at the metric \eq{eml-metric} to identify the origin of the Minkowski spacetime.
Definitely the Lorentzian manifold $\mathcal{M}$ becomes the Minkowski spacetime
when all fluctuations die out, i.e.,  $v_b^a \to \delta^a_b, \; \mathbf{A}^a \to 0$ (and so $\lambda \to 1)$.
Therefore the vacuum geometry for the metric \eq{eml-metric} was originated from the vacuum
configuration \eq{1-vaccum}.
In other words, the $(2n+1)$-dimensional Minkowski spacetime emerges from the vacuum condensate \eq{1-vaccum}
since the corresponding vielbeins and the metric are given by $E^{(0)}_A = V^{(0)}_A
= \Big( \frac{\partial}{\partial t}, \frac{\partial}{\partial y^a} \Big)$
and $ds^2 = -dt^2 + d \mathbf{y} \cdot d \mathbf{y}$ \cite{1hsy-jhep09,1hsy-review}.
The Minkowski spacetime is originated from a coherent vacuum satisfying the Moyal-Heisenberg algebra \eq{moyal-vac}. 
And the condensate \eq{vaccum} in the NC Coulomb vacuum induces a nontrivial vacuum energy density. 
We can calculate it using the action \eq{bfss-u1}:
\begin{equation}\label{1vac-energy}
    \rho_{\mathrm{vac}} = \frac{1}{4G_{YM}^2} |B_{ab}|^2.
\end{equation}
A striking fact is that the vacuum responsible for the generation of
flat spacetime is not empty. Rather the flat spacetime had been originated from the uniform
vacuum energy \eq{1vac-energy} known as the cosmological constant in general relativity.
This is a tangible difference from Einstein gravity, since Eq. \eq{1einstein-eq} enforces $T_{\mu\nu} = 0$ 
for the flat spacetime.
Consequently, the emergent gravity reveals a remarkable picture that
a uniform vacuum energy such as Eq. \eq{1vac-energy} does not gravitate (i.e., does not couple to gravity).
As a result, the emergent gravity presents a striking contrast to general relativity.
This important conclusion may be strengthened by applying the Lie algebra homomorphism \eq{lie-homo}
to the commutators in Eq. \eq{1-commrel}, which reads as
\begin{equation}\label{1lie-cosm}
-i \mathrm{ad}_{[\phi_a, \phi_b]}
\equiv \widehat{V}_{\widehat{F}_{ab}-B_{ab}}
= \widehat{V}_{\widehat{F}_{ab}} = [\widehat{V}_{a}, \widehat{V}_{b}]
 \in \mathfrak{D}^1
\end{equation}
for a constant field strength $B_{ab}$. To stress clearly, the gravitational fields emergent
from NC $U(1)$ gauge fields must be insensitive to the constant vacuum energy such as Eq. \eq{1vac-energy}.
In the end, the emergent gravity clearly dismisses the notorious
cosmological constant problem \cite{1hsy-jpcs12,1hsy-jhep09,1hsy-review}.

We observed that the MQM admits a global automorphism given by Eq. \eq{1poincare-auto}.
Let us see what is the consequence of the Poincar\'e automorphism \eq{1poincare-auto}
on the emergent spacetime geometry. The Poincar\'e automorphism leads to the transformation
$V_A^{(0)} \to V_A^{'(0)} = {\Lambda_A}^B V_B^{(0)}$. However, this transformation does not change $\lambda^2$
because $\det \Lambda = 1$. The geometry for the transformed vacuum $p'_A$ is determined
by the metric \eq{eml-metric} that is still the Minkowski spacetime $\mathbb{R}^{2n, 1}$.
Therefore, we see that the vacuum responsible for the generation of flat spacetime
is not unique but degenerate up to the Poincar\'e automorphism.\footnote{\label{1scaling-symm}Note that
the vacuum solution \eq{vaccum} is further degenerated under the scaling $p_a \to p'_a = \beta p_a$
or $y^a \to y'^a = \beta^{-1} y^a$ as far as $\beta \in \mathbb{R} \setminus \{ 0 \}$ is a nonzero constant.
We will use this freedom to normalize the initial length scale such that $|y^a(t=0)| = L_P $
or $l_s = \sqrt{\alpha'}$.}
After all, the global Poincar\'e symmetry of the Minkowski spacetime is emergent from
the Poincar\'e automorphism \eq{1poincare-auto} of MQM.

It should be remarked that the background-independent theory does not mean that the physics is independent
of the background. Background independence here means that, although a physical phenomenon
occurs in a particular background with a specific initial condition,
an underlying theory itself describing such a physical
event should presuppose neither any kind of spacetime nor material backgrounds. Therefore the
background itself should arise from a vacuum solution of the underlying theory. In particular, the
background-independent theory must integrate geometry and matter, as the matter cannot be defined 
without a pre-established spacetime framework. 
Complex spacetime structures are derived through the general deformations of the fundamental vacuum. 
These deformations correspond to physical processes that happen upon a particular (spacetime) background. 
Hence they are regarded as a dynamical system. Motivated by a close analogy with quantum mechanics,
we argued that the deformations of spacetime structure supported
on a vacuum solution must be understood as the time evolution of the dynamical system.
According to this picture, the fundamental action \eq{1mqm} describes a dynamical system,
from which an emergent $(2n + 1)$-dimensional Lorentzian spacetime $\mathcal{M}$
with the metric \eq{eml-metric} is derived.

Note that the Newton constant $G_N$ according to emergent gravity picture has to be determined
by field theory parameters only such as the gauge coupling constant $G_{YM}$
and $\theta = B^{-1}$ defining the NC $U(1)$ gauge theory.
In order to estimate the dynamical energy scale for the vacuum condensate \eq{vaccum}, 
consider a simple dimensional analysis leading to the result \cite{1hsy-jhep09,1hsy-review}
\begin{equation}\label{1gg-rel}
    \frac{G_N \hbar^2}{c^2} \sim G_{YM}^2 |\theta|,
\end{equation}
where $|\theta| := |\mathrm{Pf} \theta|^{\frac{1}{n}}$.
To be specific, when considering the four-dimensional case in which
$M_P = (8 \pi G_N)^{-1/2} \sim 10^{18}$ GeV and $G_{YM}^2 \sim \frac{1}{137}$,
the vacuum energy \eq{1vac-energy} due to the condensate \eq{vaccum} is roughly given by
\begin{equation}\label{1vac-en4}
    \rho_{\mathrm{vac}} = \frac{1}{4G_{YM}^2} |B_{ab}|^2 \sim G_{YM}^2 M_P^4 \sim 10^{-2} M_P^4.
\end{equation}
Of course, its precise value may be given when the NC scale $|\theta| = M_{nc}^{-2}$ is known.
In Eq. \eq{1vac-en4}, we roughly identified the NC scale $M_{nc}$ with the Planck energy $M_P$.
But this order of estimate is not so bad when we compare the value with that in Ref. \cite{liddle}
(see the last paragraph in Section III):
$\rho_{\mathrm{vac}}^{1/4} = V_{60}^{1/4} = 3.8 \times 10^{16}$ GeV for 60 e-foldings.
Then the inflationary Hubble parameter corresponds to $H_{60} = 2.9 \times 10^{-5} M_P$. 
Emergent gravity reveals that the enigmatic vacuum energy $\rho_{\mathrm{vac}} \sim  M_P^4$, 
rather surprisingly, serves as the true origin of flat spacetime.
If spacetime geometry emerges from a vacuum configuration of some fundamental ingredients
in an underlying quantum gravity theory, 
the Planck mass $ M_P$ is a natural dynamical scale for the emergence of
gravity and spacetime. Therefore it may not be a surprising result but rather
an inevitable consequence that the Planck energy density \eq{1vac-en4}
in vacuum was the genetic origin of spacetime.

The metric \eq{eml-metric} clearly indicates that the Planck energy condensate in vacuum resulted 
in an extremely extended spacetime. Since we have started with
a background-independent theory in which any spacetime structure has not been assumed in advance,
the spacetime was not existent at the beginning but simply emergent from the vacuum condensate \eq{vaccum}.
Therefore the Planck energy condensation into vacuum must be regarded as a dynamical process.
Since the dynamical scale for the vacuum condensate is about of the Planck energy, the time scale
for the condensation will be roughly of the Planck time $t_P \sim 10^{-44}$ sec.
Inflation scenario asserts that our Universe at the beginning had undergone an explosive inflation era
lasted roughly $\sim 10^{-33}$ seconds. Thus it is natural to consider the cosmic inflation as
a dynamical process for the instantaneous condensation of vacuum energy 
to enormously spread out spacetime \cite{1hsy-jpcs12}. Now we will explore how
the cosmic inflation is triggered by the condensate of Planck energy in vacuum
and corresponds to the dynamical emergence of spacetime.

First let us understand intuitively Eqs. \eq{1friedmann} and \eq{1inflaton-eq} to get some dear insight
from the old wisdom (see I.1 in \cite{zel-nov-book}). 
Suppose that a test particle with mass $m$ is placed in the condensate with the energy density \eq{1vac-en4}.
Consider a ball of radius $r(t)$ and the test particle placed on its surface.
According to the Gauss's law, the particle will be subject to the gravitational potential energy
$V(r) = - \frac{G_N M(r) m}{r}$ caused by the condensate \eq{1vac-en4}, where
$M (r) = \frac{4 \pi r(t)^3 \rho_{\mathrm{vac}}}{3}$ is the total mass inside the ball.\footnote{This experiment 
is a simple twist of the well-known solution of Gauss's law for gravity
inside the earth, in which the minus sign in the gravitational potential energy presupposes a repulsive force
rather than the usual attractive force. The repulsive force  in Newtonian gravity is given by $\mathbf{F} =
k_g \mathbf{r} = - \nabla V(r)$ where $k_g = \frac{4 \pi G_N m \rho_{\mathrm{vac}}}{3}$
and $V(r) = - \frac{G_N M(r) m}{2 r}$ is the gravitational potential energy.
Note that the change of sign and the factor 2 enhancement are due to the general relativity effect
since $\frac{\ddot{a}}{a} = - \frac{4 \pi G_N}{3} (\rho_{\mathrm{vac}} + 3p)
= - \frac{4 \pi G_N}{3} (- 2\rho_{\mathrm{vac}})$.}
In order to preserve the total energy $E$ of the particle, the ball has to expand so that
the kinetic energy $K(r) = \frac{1}{2} m \dot{r}(t)^2$ generated by the expansion compensates
the negative potential energy. That is, the energy conservation implies the following relation
\begin{equation}\label{1hubble}
    H^2 = \frac{8 \pi G_N \rho_{\mathrm{vac}}}{3} - \frac{k}{r(t)^2},
\end{equation}
where $H = \frac{\dot{r}(t)}{r(t)}$ is the expansion rate and $k \equiv - \frac{2E}{m}$.
Comparing the above equation with the Friedmann equation \eq{1friedmann} after the identification
$r(t) = R a(t)$, we see that Eq. \eq{1hubble} corresponds to $\rho_{\mathrm{vac}} = V(\phi) \approx V_0$ and
$\dot{\phi} \approx 0$ with $k = 0$. We actually assumed the spatially flat universe, $k = 0$,
for the Friedmann equation \eq{1friedmann}.
In our approach with a background-independent theory, the condition $k = 0$ is automatic since
the very beginning should be absolutely nothing!
This conclusion is consistent with the metric \eq{eml-metric} which describes a final state
of cosmic inflation. Hence we may claim that the background-independent theory for
cosmic inflation predicts a spatially flat universe, in which the constant $k$ must be exactly zero.

From the above simple argument \eq{1hubble} with $k=0$, 
we see that the size of the ball exponentially expands, i.e.,
\begin{equation} \label{1ball-exp}
    a(t) = a_0 e^{Ht}
\end{equation}
where
\begin{equation} \label{1hubb-h}
    H = \sqrt{\frac{8 \pi G_N \rho_{\mathrm{vac}}}{3}}
\end{equation}
is a constant. Let us introduce fluctuations around the inflating solution \eq{1ball-exp} by
considering $\rho_{\mathrm{vac}} \to \rho_{\mathrm{vac}} + \delta \rho$ and $\dot{\phi} \neq 0$,
where $\delta \rho$ is the mechanical energy due to the fluctuations of the inflaton $\phi(t)$.
Then Eq. \eq{1hubble} is replaced by
\begin{equation}\label{1hubble-fl}
    H^2 = \frac{8 \pi G_N}{3} (\rho_{\mathrm{vac}} + \delta \rho),
\end{equation}
and the dynamics of the inflaton is described by Eq. \eq{1inflaton-eq}.
The argument leading to Eq. \eq{1hubble-fl} implies that the cosmic inflation corresponds 
to a dynamical process of the Planck energy condensation into vacuum. 
Hence the cosmic inflation as a dynamical system is typically a time-dependent solution 
and must be described by a non-Hamiltonian dynamics, as we already remarked in Eq. \eq{1friction}. 
Now we will illuminate how the cosmic inflation can be described by the
conformal Hamiltonian dynamics \cite{libermarle,mcla-perl} which appears in, for example,
simple mechanical systems with friction. In Appendix A we briefly review generalized
symplectic manifolds that correspond to a natural phase space describing the conformal
Hamiltonian dynamics.

Let us consider the simplest case, namely when the symplectic manifold is $\mathbb{R}^{2n}$
with coordinates $(q^i, p_i)$ and $\omega =  dq^i \wedge dp_i = da$
where $ a=\frac{1}{2}  (q^i dp_i -  p_i dq^i)$. 
A conformal vector field $X$ is defined by
\begin{equation}\label{conf-def}
    \iota_{X} \omega = \kappa a + dH,
\end{equation}
where $H: \mathbb{R}^{2n} \to \mathbb{R}$ is the Hamiltonian and $\kappa$ is a nonzero constant.
Note that Eq. \eq{conf-def} implies
\begin{equation}\label{conformal}
    \mathcal{L}_{X} \omega = \kappa \omega.
\end{equation}
Therefore the vector field $X$ is a Lie algebra generator of conformal infinitesimal
transformations. 
It is easy to solve Eq. \eq{conf-def} for the vector field $X$ and the result is given by
\begin{equation}\label{conf-vec}
   X = \frac{\kappa}{2} \Big( q^i \frac{\partial}{\partial q^i}
   + p_i \frac{\partial}{\partial p_i} \Big) + X_H,
\end{equation}
where $X_H$ is a usual Hamiltonian vector field obeying $\iota_{X_H} \omega = dH$.
Thus the Hamilton's equations are given by
\begin{eqnarray} \label{heq-q}
&& \frac{dq^i}{dt} = X (q^i) = \frac{\kappa}{2} q^i + \frac{\partial H}{\partial p_i}, \\
  \label{heq-p}
&& \frac{dp_i}{dt} = X (p_i) = \frac{\kappa}{2} p_i - \frac{\partial H}{\partial q^i}.
\end{eqnarray}
The equations of motion for the Hamiltonian $H = \frac{1}{2} p_i^2 + U(q)$ are equal to 
the differential equations
\begin{equation}\label{friction-deq}
\ddot{q}^i - \kappa \dot{q}^i + \frac{\partial V}{\partial q^i} = 0,
\end{equation}
where $V(q) = U(q) + \frac{\kappa^2}{8} q_i^2$.
To be specific, the integral curves for $U(q) = \frac{1}{2} \omega^2 q_i^2$ are given by\footnote{Note that
$a = b + d \lambda$ where $b = - p_i dq^i$ and $\lambda = \frac{1}{2} q^i p_i$. Thus one can also define
the conformal vector field $X$ by $\iota_{X} \omega = \kappa b + dH'$
where $H' = H + \kappa\lambda$. In this case $X = \kappa p_i \frac{\partial}{\partial p_i}
+ X_{H'}$ and the equations of motion are given by $\frac{dq^i}{dt} = \frac{\partial H'}{\partial p_i}$
and $\frac{dp_i}{dt} = \kappa p_i - \frac{\partial H'}{\partial q^i}$. For $H' = \frac{1}{2}
(p_i^2 + \omega^2 q_i^2)$, the general solution is $q^i (t) = A^i e^{\frac{\kappa}{2} t}
\sin \Big(\sqrt{\omega^2 - \frac{\kappa^2}{4}} t + \theta \Big)$, 
which describes a damped harmonic oscillator when $\kappa <0$. However the vector field defined
by Eq. \eq{conf-vec} is more convenient for our purpose.}
\begin{equation}\label{eom-sol}
    q^i (t) =  e^{\frac{\kappa}{2} t} q^i(\kappa = 0; t), \qquad
    p_i (t) =  e^{\frac{\kappa}{2} t} p_i(\kappa = 0; t),
\end{equation}
where $q^i(\kappa = 0; t) = A^i \sin (\omega t + \theta)$ and
$p_i(\kappa = 0; t) = B_i \cos (\omega t + \theta)$ describe the usual harmonic oscillator
with a closed orbit when $\kappa = 0$.
The flow generated by a conformal vector field can be directly obtained by integrating Eq. \eq{conformal}. 
Let $\phi_t$ denote the flow of $X$. By the Lie derivative theorem \cite{sg-book},
we have $\frac{d}{dt} (\phi_t^* \omega) = \phi_t^* \mathcal{L}_X \omega = \kappa \phi_t^* \omega$.
Therefore we see that the conformal flow has the property
\begin{equation}\label{time-omega}
    \phi^* \omega =  e^{\kappa t} \omega.
\end{equation}
This means that the volume of phase space exponentially expands (contracts)
if $\kappa > 0 \; (\kappa < 0)$.

The mathematical parallel between quantum mechanics and NC spacetime offers insights 
into formulating cosmic inflation as a dynamical system.
First note that the NC space \eq{1extra-nc2n} in commutative limit becomes a phase space
with the symplectic form
\begin{equation}\label{const-b}
    B = \frac{1}{2} B_{\mu\nu} dy^\mu \wedge dy^\nu.
\end{equation}
Hamiltonian systems generated by divergenceless Hamiltonian flows are characterized 
by the invariance of phase space volume under time evolution, 
which is known as the Liouville theorem \cite{sg-book}.
However, the cosmic inflation indicates that the volume of spacetime phase space
has to exponentially expand as seen from the above mechanical analogue.
Hence a generalized Liouville theorem is necessary to describe the exponential expansion of spacetime.
We have already observed how a non-Hamiltonian dynamics can be formulated
in terms of a {\it conformal} Hamiltonian dynamics characterized by the flow obeying Eq. \eq{conformal}. 
See Appendix A for a mathematical exposition of general
time-dependent nonconservative dynamical systems.

Let us apply the conformal Hamiltonian dynamics to the cosmic inflation.
Recall that we have considered an atlas $\{(U_i, \phi_{(i)}) \}$ on $M = \bigcup_{i \in I} U_i$
as a collection of local Darboux charts and complete it by gluing these local charts on their overlap.
On each local chart, we have a local symplectic structure $\Omega_i = \frac{1}{2} B_{\mu\nu} dy^\mu_{(i)}
\wedge dy^\nu_{(i)}$ where $\{y^\mu_{(i)} \}$ are Darboux coordinates on a local patch $U_i \subset M$.
The phase space coordinates $\{y^\mu_{(i)} \}_{U_i}$ of a conformal Hamiltonian system undergo a nontrivial
time evolution even in a local Darboux frame \cite{vaisman2,clm-carolin}. For example, the equations of
motion \eq{heq-q} and \eq{heq-p} illustrate such a nontrivial time evolution even when $H=0$.
The dynamics in this case consists of the orbits of a conformal vector field $X$ obeying
the condition \eq{conformal}. The situation at hand is essentially the same as the mechanical
system with negative-friction. To be specific, write $\Omega_i = d a_{(i)}$ on a local patch $U_i \subset M$
where $a_{(i)} = - \frac{1}{2} p^{(i)}_\mu dy^\mu_{(i)}$ and $p^{(i)}_\mu = B_{\mu\nu} y^\nu_{(i)}$. 
Define a conformal vector field $X$ as 
\begin{equation}\label{inf-conf}
    \iota_{X} \Omega_i = \kappa a_{(i)} + dH_i,
\end{equation}
where $H_i: U_i \to \mathbb{R}$ is a local Hamiltonian and $\kappa$ is a positive constant.
Using the fact that $d \Omega_i = 0$, Eq. \eq{inf-conf} can be written as
\begin{equation}\label{inf-conformal}
    \mathcal{L}_{X} \Omega_i = \kappa \Omega_i.
\end{equation}
The vector field $X$ obeying Eq. \eq{inf-conf} is given by
\begin{equation}\label{inf-vec}
   X = \frac{\kappa}{2} y^\mu_{(i)} \frac{\partial}{\partial y^\mu_{(i)}} + X_{H_i},
\end{equation}
where $X_{H_i}$ is the ordinary Hamiltonian vector field satisfying $\iota(X_{H_i}) \Omega_i = dH_i$.
The conformal vector field \eq{inf-vec} contains the Liouville vector field 
$Z_{(i)} \equiv \frac{1}{2} y^\mu_{(i)} \frac{\partial}{\partial y^\mu_{(i)}}$ \cite{libermarle,mcla-perl}.

Let us consider a spacetime dynamics generated by the Liouville vector field.
We will set $H_i=0$ for simplicity.
The time evolution of local Darboux coordinates is then determined by the equations
\begin{equation}\label{time-darboux}
    \frac{d y^\mu_{(i)}}{dt} = X(y^\mu_{(i)}) = \frac{\kappa}{2} y^\mu_{(i)}.
\end{equation}
The solution is given by
\begin{equation}\label{exp-sol}
    y^\mu_{(i)} (t) =  e^{\frac{\kappa}{2} t} y^\mu_{(i)}(0).
\end{equation}
We may glue the local solutions \eq{exp-sol} to have a global form
\begin{equation}\label{exp-gsol}
    p_a (t) = B_{ab} y^b (t) = e^{\frac{\kappa}{2} t} p_a.
\end{equation}
Then the time-dependent canonical one-form is given by
\begin{equation}\label{tcan-1}
    a(t) = - \frac{1}{2} p_a (t) dy^a (t) = - \frac{1}{2} e^{\kappa t} p_a dy^a
\end{equation}
and thus
\begin{equation}\label{t-symp}
    \Omega (t) = d a(t) = e^{\kappa t} B.
\end{equation}
The exterior derivative above acts only on $\mathbb{R}^{2n}$.
One can show that the result \eq{t-symp} is the integral form of Eq. \eq{inf-conformal}. 
More generally, the result \eq{t-symp} is a particular case of
the general Moser flow $\phi_t$ generated by a time-dependent vector field $X_t$
for a locally conformal symplectic manifold \cite{bako}. The Moser flow satisfies 
\begin{equation}\label{moser-flow}
    \phi_t^* \Omega_t = \exp \Big( \int_0^t \phi_s^* \big( b_s(X_s) \big)  ds \Big) \cdot \Omega,
\end{equation}
where the one-form $b$ is the Lee form of $\Omega$ \cite{hclee}.
The result \eq{t-symp} is simply obtained from Eq. \eq{moser-flow} when $b(X)$ is a constant $\kappa$.

We have advanced the concept of cosmic inflation by postulating that the vacuum configuration \eq{1-vaccum} 
serves as a final state accumulating the vacuum energy.
Therefore, the cosmic inflation corresponds to a dynamical system describing
the transition from the initial state referring to ``absolutely nothing" to the final state.
With this perspective in mind, let us consider a symplectic manifold $\big(M,  \Omega (t) \big)$
whose symplectic two-form is given by Eq. \eq{t-symp}.
It can be shown that this symplectic manifold arises from a time-dependent vacuum
given by
\begin{equation}\label{t-vacuum}
    \langle \phi_a (t) \rangle_{\mathrm{vac}} =  p_a (t) = e^{\frac{\kappa}{2} t} p_a,
    \qquad \langle A_0 (t) \rangle_{\mathrm{vac}} = 0.
\end{equation}
Recall that the temporal gauge field in Eq. \eq{t-vacuum} corresponds to our previous setting $H_i =0$ 
according to the identification \eq{local-ham}.
Though we have turned off the temporal gauge field for a simple argument, 
it is necessary to turn on it in order to implement the vacuum \eq{t-vacuum} 
as a solution of the action \eq{1mqm}. We will consider it later. 
Let us first determine the vacuum geometry emergent from the vacuum configuration \eq{t-vacuum}.
Note that
\begin{equation}\label{t-vac-sym}
    \langle [\phi_a (t), \phi_b (t)] \rangle_{\mathrm{vac}} = - i e^{\kappa t} B_{ab} = - i \Omega_{ab} (t),
\end{equation}
and so we regard $\Omega (t) = \frac{1}{2} \Omega_{ab} (t) dy^a \wedge dy^b$ as the symplectic structure
of the inflating vacuum \eq{t-vacuum}.
The vacuum \eq{t-vacuum} leads to the vector fields (omitting the symbol indicating the vacuum
for a notational simplicity)
\begin{equation}\label{vac-infvecs}
    V_A (t) = \left(V_0 (t), V_a (t) \right)  = \left(\frac{\partial}{\partial t}, e^{\frac{\kappa}{2} t} V_a (0) \right),
\end{equation}
where $V_a (0) = \delta^\mu_a \frac{\partial}{\partial y^\mu}$.
Thus the dual one-forms are given by
\begin{equation}\label{t-1form}
    v^0(t) = dt, \qquad v^a (t) = e^{- \frac{\kappa}{2} t} v^a (0)
\end{equation}
where $v^a (0) = \delta^a_\mu dy^\mu$.
It is easy to calculate the Lie algebra \eq{vstr-eq}
for the time-dependent vector fields $V_A (t)$: 
\begin{equation}\label{g-desitter}
    {g_{AB}}^C = \left\{
                   \begin{array}{ll}
                     {g_{0a}}^b = - {g_{a0}}^b = - \frac{\kappa}{2} \delta^b_a,  & a, b = 1, \cdots, 2n; \\
                     0, & \hbox{otherwise.}
                   \end{array}
                 \right.
\end{equation}
Thus $\lambda^2 =  e^{ n \kappa t}$ according to Eq. \eq{vol-lamb}.
The invariant volume form of the vacuum manifold is then given by
\begin{equation}\label{t-vol}
    \nu_t =  \lambda^2 dt \wedge  v^1(t) \wedge \cdots  \wedge v^{2n}(t)
    = dt \wedge  dy^1 \wedge \cdots  \wedge dy^{2n}.
\end{equation}
After applying the above results to the metric \eq{eml-metric}, we see that
the vacuum configuration \eq{t-vacuum} determines the spacetime geometry with the metric
\begin{equation}\label{ds-metric}
    ds^2 = - dt^2 + e^{2 H t} d\mathbf{y} \cdot d\mathbf{y}
\end{equation}
where $H = \frac{1}{2} (n-1) \kappa > 0$ is a positive Hubble parameter. 
This is the de Sitter space in flat coordinates which covers half of the de Sitter manifold.
Definitely the inflation metric \eq{ds-metric} describes a homogeneous and isotropic Universe
known as the Friedmann-Robertson-Walker metric in physical cosmology.

The vector fields $V_A (t)$ in Eq. \eq{g-desitter} form a solvable Lie algebra and the de Sitter space is its Lie group.
The Lie algebra has the generators $V_0 = -\frac{\kappa}{2} L_{0(2n+1)}, \;
V_a = \frac{1}{2} (L_{0a} + L_{a(2n+1)})$, which is indeed a subalgebra of the de Sitter algebra
where $L_{\mathbb{A}\mathbb{B}} \; (\mathbb{A}, \mathbb{B} = 0,1, \cdots, 2n+1)$ are the Lie algebra generators of $SO(2n + 1, 1)$ Lorentz symmetry.
In this point of view, energy and momentum do not commute unlike in the Minkowski spacetime
and are no longer conserved, as translations are no more a symmetry of
the space.\footnote{One important consequence is that the energy will not be positive.
Polyakov has suggested \cite{polya} that this makes de
Sitter space unstable with respect to decay by creation of particle-antiparticle pairs.}
Instead, energy generates scale transformations in momentum. This is the reason
why the isometry of the de Sitter space is enhanced to $SO(2n+1, 1)$ which combines $SO(2n,1)$
Lorentz transformations and translations together \cite{hawk-elli}. In the limit $\kappa \to 0$,
we recover the Minkowski spacetime.

In order to achieve a background-independent formulation of emergent spacetime, 
it is desirable to realize the inflationary universe
as a solution of the matrix model \eq{1mqm}. Now we will show that the cosmic inflation
arises as such a time-dependent solution describing the dynamical process of Planck energy condensate
into vacuum without introducing any inflaton field as well as an {\it ad hoc} inflation potential.
First let us show that the dynamical process for the vacuum condensate
is described by the time-dependent vacuum configuration given by
\begin{equation}\label{1time-vacuum}
    \langle \phi_a (t) \rangle_{\mathrm{vac}} = p_a (t) = e^{\frac{\kappa t}{2}} p_a, \qquad
    \langle \widehat{A}_0 (t) \rangle_{\mathrm{vac}} = \widehat{a}_0 (t, y),
\end{equation}
where the temporal gauge field is given by an open Wilson line \cite{1open-wilson3}
\begin{equation}\label{1open-wilson}
 \widehat{a}_0 (t, y) = \frac{\kappa}{2} \int_0^1 d \sigma \frac{d y^a (\sigma)}{d \sigma} p_a (\sigma)
\end{equation}
along a path parameterized by the curve $y^a (\sigma) = y_0^a + \zeta^a (\sigma)$
where $\zeta^a (\sigma) = \theta^{ab} k_b \sigma$ with $0 \leq \sigma \leq 1$
and $y^a (\sigma=0) \equiv y_0^a$ and $y^a (\sigma=1) \equiv y^a$.
The constant $\kappa$ will be identified with the inflationary Hubble constant $H$.
Note that the second term in Eq. \eq{eom-mqm} identically vanishes
for the background \eq{1time-vacuum}. Therefore it is enough to impose the condition
\begin{equation}\label{1back-sol}
     D_0 \phi_a = e^{\frac{\kappa t}{2}} \Big( \frac{\kappa}{2} p_a - i [\widehat{A}_0, p_a] \Big) = 0
\end{equation}
to satisfy both \eq{eom-mqm} and \eq{gauss-mqm}.
In terms of the NC $\star$-algebra $\mathcal{A}_\theta^1$, Eq. \eq{1back-sol} reads as
\begin{equation}\label{1tback-eq}
    \frac{\partial \widehat{a}_0 (t, y)}{\partial y^a} = \frac{\kappa}{2} p_a.
\end{equation}
Using the formula
\begin{equation}\label{1wilson-formula}
    \frac{\partial}{\partial y^a} \int_0^1 d\sigma \frac{dy^b (\sigma)}{d \sigma}
K \big( y(\sigma) \big) = \delta^b_a K(y)
\end{equation}
for some differentiable function $K(y)$, one can easily check that the temporal gauge field
in Eq. \eq{1open-wilson} satisfies Eq. \eq{1tback-eq}.

We want to address some physical significance of the nonlocal term \eq{1open-wilson}.
It is essential to highlight that the temporal gauge field \eq{1open-wilson} corresponds to
a background Hamiltonian density in the comoving frame.\footnote{\label{1comove}This feature 
is due to our choice of a coordinate frame to describe the dynamical system. The time evolution operator $\widehat{\phi}_0 (t, y) = i \frac{\partial}{\partial t} + \widehat{A}_0 (t, y)$ is defined in the comoving frame. 
In general, one may choose an arbitrary frame in which
the time evolution is described by $k(t, y) \frac{\partial}{\partial t}
\in \mathrm{Hor}(\mathcal{A}^1_\theta)$.
A particularly interesting frame is the conformal coordinates with which 
the metric is given by $ds^2
= a(\eta)^2 (-d\eta^2 + d\mathbf{x} \cdot d\mathbf{x})$ where $a(\eta) = - \frac{1}{H \eta}$
and $-\infty < \eta < 0$. The conformal coordinates can be easily transformed
to the comoving coordinates by $a(\eta) d\eta = dt$.}
It will be shown that, although the temporal gauge field \eq{1open-wilson} is nonlocal, 
the gravitational metric determined by the time-dependent vacuum configuration \eq{1time-vacuum} 
still takes a local expression as it should be. 
It was already noticed in \cite{1lry-jhep} that nonlocal observables
in emergent gravity are in general necessary to describe some gravitational metric. 
Indeed the appearance of such nonlocal terms should not be surprising 
since there exist no local gauge invariant observables in NC gauge theories \cite{1open-wilson3}.

Now let us determine the metric \eq{eml-metric} for the inflating background \eq{1time-vacuum}.
The $(2n+1)$-dimensional vector fields defined by Eq. \eq{time-derivation} take the following form
\begin{equation}\label{1vac-vec}
    V_0 (t) = \frac{\partial}{\partial t} - \frac{\kappa}{2} y^a \frac{\partial}{\partial y^a},
    \qquad V_a (t) = e^{\frac{\kappa t}{2}} \frac{\partial}{\partial y^a}.
\end{equation}
Higher-order derivative terms in Eqs. \eq{time-polyvec} and \eq{space-polyvec} identically vanish 
since only the vacuum background \eq{1time-vacuum} was considered. 
Note that the vector fields take the local form again
as the result of applying the formula \eq{1wilson-formula} and the open Wilson line \eq{1open-wilson}
leads to a conformal vector field $Z \equiv \frac{1}{2} y^a \frac{\partial}{\partial y^a}$
known as the Liouville vector field \cite{libermarle,mcla-perl}.
Then the dual orthogonal one-forms are given by
\begin{equation}\label{1vac-dual}
    v^0 (t) = dt, \qquad v^a (t) = e^{-\frac{\kappa t}{2}} (dy^a + \mathbf{a}^a) = e^{- \kappa t} dy_t^a
\end{equation}
where
\begin{equation}\label{1d-mod-a0}
\mathbf{a}^a = \frac{\kappa}{2} y^a dt, \qquad y^a_t \equiv  e^{\frac{\kappa t}{2}} y^a.
\end{equation}
One can see that the vector fields in Eq. \eq{1vac-vec} satisfy $[V_0, V_a]  =  \kappa V_a$
and thus
\begin{equation}\label{1g-desitter}
    {g_{AB}}^C = \left\{
                   \begin{array}{ll}
                     {g_{0a}}^b = - {g_{a0}}^b = - \kappa \delta^b_a,  & a, b = 1, \cdots, 2n; \\
                     0, & \hbox{otherwise.}
                   \end{array}
                 \right.
\end{equation}
From this result, we get $\lambda^2 =  e^{2 n \kappa t}$ since ${g_{B A}}^B = V_A \ln \lambda^2$ \cite{1hsy-review,ly-jkps}.
One can see that the volume-preserving condition \eq{vol-new} is satisfied since
$\rho = e^{n \kappa t}, \; A_0^a = - \frac{\kappa}{2} y^a$ and $V_a^\mu = e^{\frac{\kappa t}{2}} \delta_a^\mu$.
In the end, the time-dependent metric for the inflating background \eq{1time-vacuum} is given by
\begin{equation}\label{1inflation}
    ds^2 = - dt^2 + e^{2Ht} d\mathbf{y}_t \cdot d\mathbf{y}_t,
\end{equation}
where we have identified the inflationary Hubble constant $H \equiv (n-1) \kappa$.
By comparing this result with Eq. \eq{ds-metric}, one can see that the temporal gauge field \eq{1open-wilson}
enhances the inflation by the factor two, i.e. $ H \to 2H$.
We emphasize that the temporal gauge field \eq{1open-wilson} is crucial to satisfy Eqs. \eq{eom-mqm}
and \eq{gauss-mqm}.

We demonstrated that cosmic inflation arises as a time-dependent solution of a background-independent theory. 
This theory delineates the dynamical evolution of the Planck energy condensate in a vacuum, 
without introducing an inflaton field or an {\it ad hoc} inflation potential.
Let us generalize the cosmic inflation by also including arbitrary fluctuations
around the inflationary background \eq{1time-vacuum}.
Such a general inflationary universe in $(2n + 1)$-dimensional Lorentzian spacetime
can be realized by considering a time-dependent NC algebra given by
\begin{equation}\label{1time-ncalg}
    {}^t\mathcal{A}_\theta^1 \equiv \Big\{ \widehat{\phi}_0 (t, y) = i \frac{\partial}{\partial t}
    + \widehat{A}_0 (t, y), \quad \widehat{\phi}_a (t, y) = e^{\frac{\kappa t}{2}} \big( p_a
    + \widehat{A}_a (t, y) \big) \Big\}.
\end{equation}
We denote the corresponding time-dependent matrix algebra by ${}^t\mathcal{A}_N^1$ which
consists of time-dependent solutions of the action \eq{1mqm}.
Then the general Lorentzian metric describing a $(2n + 1)$-dimensional inflationary universe
can be obtained by the following duality chain:
\begin{equation}\label{1inf-chain}
  {}^t\mathcal{A}^1_N  \quad \Longrightarrow \quad {}^t\mathcal{A}^1_\theta \quad
  \Longrightarrow \quad {}^t\mathfrak{D}^1.
\end{equation}
The module ${}^t\mathfrak{D}^1$ of derivations of the NC algebra ${}^t\mathcal{A}_\theta^1$ is
given by
\begin{equation}\label{1inf-der}
    {}^t\mathfrak{D}^1 = \Big\{ \widehat{V}_A (t) = (\widehat{V}_0, \widehat{V}_a) (t) |
    \widehat{V}_0 (t) = \frac{\partial}{\partial t} + \mathrm{ad}_{\widehat{A}_0 (t, y)},
    \quad \widehat{V}_a (t) = e^{\frac{\kappa t}{2}} \Big( \frac{\partial}{\partial y^a} +
    \mathrm{ad}_{\widehat{A}_a (t, y)} \Big) \Big\},
\end{equation}
where the adjoint operations are defined by Eq. \eq{inner-d}.
In the classical limit of the module \eq{1inf-der}, we get a general inflationary universe described by
\begin{equation}\label{1gen-inf}
    ds^2 = -dt^2 + e^{2Ht} (1 + \delta \lambda)^2 v^a_b v^a_c (dy_t^b - \mathbf{A}^b) (dy_t^c - \mathbf{A}^c),
\end{equation}
where $v^a_b:= v^a_b(t, y), \; \delta\lambda := \delta \lambda (t, y)$ and $\mathbf{A}^b := \delta a_0^b (t, y) dt$.
If all fluctuations are turned off for which $v^a_b = \delta^a_b$ and $\delta \lambda = \mathbf{A}^b = 0$,
we recover the inflation metric \eq{1inflation}.

Let us bring this section to a close by delving into the physical implications 
arising from the results we have garnered.
Recall that a NC space such as $\mathbb{R}^2_{\theta}$ does not admit a state 
defined on a single point of the plane but the state rather lies in a region of the plane.
Thus there must be a basic length scale, below which the notion of space (and time)
does not make sense. Let us fix such a typical length scale at $t = 0$
as $|y^a(t = 0)| \sim L_P$ or $l_s = \sqrt{\alpha'}$ using the scaling freedom noted
in footnote \ref{1scaling-symm}.
Since we have started with a background-independent theory in which
a spacetime structure has to be created from a solution at the beginning $t = 0$, 
it should be reasonable to identify $L_P$ with the Planck length.
Since $y^a(t = 0)$ are operators acting on a Hilbert space, this means that
the inflationary vacuum \eq{1time-vacuum} creates a spacetime of the Planck size.
After the creation, the universe undergoes the inflation epoch described by a solution of
time-dependent matrix model unlike the traditional inflationary models that suppose 
just the exponential expansion of a preexisting spacetime.
This picture is similar to the birth of inflationary universes in Ref. \cite{1hmv} in which
the universe is spontaneously created by quantum tunneling from nothing into a de Sitter space.
Here, the "nothing" means a state devoid of any spacetime structure.
According to the standard inflation scenario, the universe expanded by at least a factor
of $e^{60} \sim 10^{26}$ during the inflation. 
After 60 e-foldings at $t = t_{\mathrm{end}} = 10^{-36} \sim 10^{-33}$ sec, 
$Ht_{\mathrm{end}} \gtrsim 60$ and the size of universe at the end of
inflation amounts to $|y^a(t = t_{\mathrm{end}})| = e^{Ht_{\mathrm{end}}} |y^a(t = 0)|
\gtrsim 10^{26} L_P$. Since 1 eV $= (6.6 \times 10^{-16} \mathrm{sec})^{-1}$,
this roughly informs us of the energy scale of the inflationary Hubble constant
$H \gtrsim 10^{11} \sim  10^{14} \mathrm{GeV}$ \cite{1guth,1linde,liddle}.

Since the vacuum \eq{1time-vacuum} is in highly non-equilibrium 
(i.e., time-dependent), 
it is expected that it undergoes evolutionary processes towards its final state \eq{1-vaccum} through interactions with its environment, such as ubiquitous fluctuations. 
This dissipation process of inflation energy is known as the reheating mechanism 
in physical cosmology. 
To accurately ascertain the duration of inflation, 
the precise mechanism involved in reheating must be understood; 
unfortunately, this surpasses our current knowledge.
Nevertheless, we will speculate in Sec. 4 a plausible picture for the reheating mechanism.

\section{Discussion}

String theory has been developed upon two distinct spacetime frameworks, namely the Kaluza-Klein (KK) theory and emergent gravity. Despite their conceptual divergence, these models represent exclusive perspectives on the nature of spacetime.
On the one hand, the KK gravity is defined in higher dimensions as a more superordinate theory and gauge theories in lower dimensions are derived from the KK theory via compactification. 
Since the KK theory is just the Einstein gravity in higher dimensions, 
the prior existence of spacetime is {\it a priori} assumed.
On the other hand, in emergent gravity picture, gravity in higher dimensions is not a fundamental force 
but a collective phenomenon emergent from more fundamental ingredients defined in lower dimensions.
In emergent gravity approach, the existence of spacetime is not {\it a priori} assumed 
but the spacetime structure is defined by the theory itself. 
This picture leads to the concept of emergent spacetime.
In some sense, emergent gravity is the inverse of KK paradigm, schematically summarized by \cite{1essay}
\begin{equation}\label{1kk-em}
    (1 \otimes 1)_S \rightleftarrows 2 \oplus 0
\end{equation}
where $\rightarrow$ means the emergent gravity picture while $\leftarrow$ indicates the KK picture.

Recent developments in string theory have revealed growing evidences for emergent gravity
and emergent spacetime. The AdS/CFT correspondence and matrix models are typical examples
supporting the emergence of gravity and spacetime \cite{1ads-cft}.
An intriguing aspect is that the emergence of gravity requires the emergence of spacetime too.
If spacetime is emergent, everything supported on the spacetime should be emergent too,
ensuring internal consistency within the theoretical framework. 
In particular, matters cannot exist without
spacetime and thus must be emergent together with the spacetime.
Eventually, the background-independent theory has to make no distinction between
geometry and matter \cite{1q-emg}.
This is the reason why the emergent spacetime picture cannot coexist peacefully with the KK paradigm.
Since the emergent spacetime is a new fundamental
paradigm for quantum gravity and radically different from any previous physical theories,
all of which describe what happens in a given spacetime,
there is a compelling need to critically reassess the underpinnings of quantum gravity 
through the lens of emergent spacetime. 
Quantum gravity is considered necessary for a complete understanding of cosmic inflation 
because inflationary theory involves the extreme conditions of the early universe 
where both quantum mechanics and gravity play significant roles.

It is well-known \cite{ncft-sw,1nc-bst,1nc-seiberg} that NC field theories arise as a low-energy effective theory in string theory, in particular, on D-branes upon turning on a constant $B$-field.
A remarkable aspect of the NC field theory is that it can be mapped to a large $N$ matrix model
as depicted in the isomorphism \eq{iso-matop}. The relation between NC gauge theories and matrix models is quite
general since any Lie algebra or Moyal-type NC space such as \eq{1extra-nc2n} always admits
a separable Hilbert space and NC gauge fields become operators acting on the Hilbert
space \cite{1hsy-epjc09}. The matrix representation of NC gauge fields implies that they can be
embedded into a background-independent formulation in terms of a matrix model.
The background-independent variables are identified as the degrees of freedom inherent in the underlying matrix model.
The relation between a matrix model and a NC gauge theory is based on the observation \cite{1hea,1hsy-epjc09} 
that the NC space \eq{1extra-nc2n}
is a consistent vacuum solution of a large $N$ gauge theory in the Coulomb branch.
The matrices are original dynamical variables of the matrix model which are manifestly
background-independent and NC gauge fields are derived from fluctuations 
in the NC Coulomb branch.

We have shown that the cosmic inflation arises as a solution of a time-dependent matrix model,
describing the dynamical process of the vacuum energy condensation. 
Remarkably, the inflation can be described by time-dependent matrices only 
without introducing any inflaton field as well as an {\it ad hoc} inflation potential. 
In order to describe the cosmic inflation, it is necessary to generalize symplectic manifolds,
as we have discussed the rationales in Sec. 3.
The corresponding generalized symplectic manifolds for the cosmic inflation include 
locally conformal symplectic (LCS) or more generally locally conformal cosymplectic (LCC) manifolds, whose mathematical foundation will be reviewed in Appendix A. 
The LCS manifold allows a nontrivial conformal vector field defined
by Eq. \eq{inf-conformal} even when an underlying Hamiltonian function identically vanishes.
The so-called Liouville vector field $Z \equiv \frac{1}{2} y^\mu \frac{\partial}{\partial y^\mu}$
is still nontrivial \cite{libermarle} and it generates the exponential expansion of spacetime described
by the metric \eq{ds-metric}.\footnote{It may be remarked that it is not possible
to realize the Liouville vector field in terms of a local Hamiltonian function.
Thus the inflation is a dynamical system without Hamiltonian. 
However we present some example in Appendix B showing that
this situation may be cured by introducing a time-dependent Hamiltonian.}
If the one-form $a$ in Eq. \eq{inf-conf} is proportional to the Lee form $b$, $X$ is
called a Hamiltonian vector field of an LCS manifold. See the definition \eq{ham-vec}.
The Hamiltonian vector field in this case shows a peculiar property different
from the symplectic case: If $b$ is not exact, $X_H = 0$ only if $H = 0$ 
(see Proposition 2.1 in \cite{vaisman2}). 
Therefore we see that the vector fields of an LCS manifold is in
stark contrast to those of a symplectic manifold, in which $X_H = 0$ if and only if 
$H =$ constant. Due to this property, while the constant vacuum energy 
(i.e., a cosmological constant) does not couple to gravity if gravity is described 
by a symplectic manifold, 
the vacuum energy rightly couples to gravity during the inflation 
if the cosmic inflation is described by an LCS (or more generally LCC) manifold.
This is a desirable property since the cosmic inflation is triggered by the condensate
of vacuum energy. Physically the reason is obvious since every quantities during the inflation
are time-dependent due to the existence of the nontrivial Liouville vector field.

It may be instructive to understand the above situation more closely in comparison
with the equilibrium case described by the metric \eq{eml-metric}.
First note that the invariant volume form \eq{vol-t} can be written as
\begin{equation}\label{2-vol}
 \nu_t = \lambda^{2-2n} \nu_g,
\end{equation}
where $\nu_g = e^0 \wedge \cdots \wedge e^{2n} = \sqrt{-\mathcal{G}} d^{2n+1} x$ is the volume form
of the metric. Therefore, the vector fields $V_A$ do not necessarily preserve the Riemannian volume
form $\nu_g$ although they preserve the volume form $\nu_t$.
However,  since $\lambda^2 \to 1$ at spatial infinity according to Eq. \eq{vol-lamb},
$\nu_t|_\infty = \nu_g|_\infty$ for the asymptotic volume forms denoted by $\nu_t|_\infty$
and $ \nu_g|_\infty$.
Therefore, the flow generated by $V_A$ leads to only local changes of the spacetime volume
while it preserves the volume element at asymptotic
regions. On the contrary, the conformal vector field \eq{inf-vec} changes the spacetime volume everywhere.
Accordingly it definitely gives rise to the exponential expansion of the spacetime volume.
After all, we see that a natural phase space for the cosmic inflation has to contain an LCS manifold instead of a standard symplectic manifold. 
Including time, it becomes an LCC manifold \cite{clm-carolin}.
Our result shows that the matrix model \eq{bfss} contains the LCC manifold as a solution.

An important question is whether the emergent spacetime picture can also lead to the eternal
(or chaotic) inflation. The answer is certainly no. The reason is the following.
We showed that the inflationary vacuum \eq{1time-vacuum} arises as a solution of
the (BFSS-like) matrix model \eq{1mqm}.
In order to define the matrix model \eq{1mqm}, however, we have not introduced any spacetime structure.
Hence the vacuum \eq{1time-vacuum} corresponds to the creation of spacetime unlike
the traditional inflationary models that describe just the
exponential expansion of a preexisting spacetime.
More precisely, the inflationary vacuum \eq{1time-vacuum} describes a dynamical process
of the Planck energy condensate responsible for the emergence of spacetime.
In general relativity, the Minkowski spacetime with the metric $g_{\mu\nu} = \eta_{\mu\nu}$
must be a completely empty space because the Einstein equation \eq{1einstein-eq} requires
$T_{\mu\nu} = 0$. However, in emergent gravity, it is not an
empty space but the vacuum condensate of the Planck energy as Eq. \eq{1vac-en4} clearly indicates.
An important point is that the Planck energy condensate results in a highly coherent vacuum
called the NC space. And the NC space is identical to the NC phase space in quantum mechanics which necessarily brings about the Heisenberg's uncertainty relation, $\Delta x \Delta p \geq \frac{\hbar}{2}$. Thus the NC space \eq{1extra-nc2n} also leads to the spacetime uncertainty relation.
Therefore any further accumulation of energy over the vacuum \eq{1time-vacuum}
must be subject to the spacetime exclusion principle known as the UV/IR mixing \cite{1uv-ir}.
Consequently, it is not possible to further accumulate the Planck energy
density over the inflationary vacuum \eq{1time-vacuum}.
This means that it is impossible to superpose a new inflating subregion over the inflationary vacuum.
Rather, it was argued \cite{ly-dedm} that the UV/IR mixing due to the spacetime uncertainty principle 
gives rise to a late-time acceleration of the universe, a.k.a. the dark energy.

In sum, the cosmic inflation triggered by the Planck energy condensate into vacuum must be
a single event \cite{1hsy-jpcs12} and the emergent spacetime precludes 
the formation of pocket universes appearing in the eternal (or chaotic) inflation. 
In the end we have a beautiful picture:
The NC spacetime is necessary for the emergence of spacetime and the exclusion principle
of NC spacetime guarantees the stability of spacetime.

We certainly live in the universe where the inflationary epoch had lasted only for a very tiny
period in very early times although it is currently in an accelerating phase driven by the dark energy.
Therefore there should be some relaxation mechanism for the (first-order) phase transition from the
inflating universe to a radiation-dominated universe. We showed that the former is described by the metric \eq{1gen-inf} whereas the latter is described by \eq{eml-metric} and
both arise as solutions of the background-independent matrix model \eq{bfss}.
In inflation scenarios in terms of scalar fields, the relaxation mechanism is known
as the reheating in which the scalar field switches from being overdamped to being underdamped and
begins to oscillate at the bottom of the potential to transfer its energy to a radiation dominated plasma
at a sufficiently high temperature to allow standard big bang nucleosynthesis \cite{zel-nov-book}.
For this purpose, most inflationary theories have introduced a very {\it ad hoc} potential 
for the scalar field (inflaton).
In our case, however, we have introduced neither an inflaton field nor an inflation potential.
Therefore, the important question is how to end the inflation of our universe 
in the emergent gravity.

We do not know the precise mechanism for the graceful exit.
Thereby we will briefly speculate a plausible scenario only.
Let us start with a naive observation.
The Lorentzian metric \eq{1gen-inf} describes general scalar-tensor perturbations
on the inflating spacetime. Since the fluctuations have been superposed on
the inflating background, we suspect that there may be some nonlinear damping mechanism
through the interactions between the background and the density fluctuations.
To be precise, there may be a cosmic analogue of the Landau damping in plasma physics
originally applied to longitudinal oscillations of an electron plasma.
The Landau damping in a plasma occurs due to the energy exchange between an electromagnetic wave
and particles in the plasma with velocity approximately equal to phase velocity of the wave.
It leads to exponentially decaying collective  oscillations.\footnote{There is a nice exposition
on the Landau damping by Werner Herr, ``Introduction to Landau Damping," available at https://cds.cern.ch/record/1507631/files/CERN-2014-009.pdf. 
Recently the Landau damping has been mathematically established even at the non-linear
level \cite{landau-vilani}.}
The Landau damping may be intuitively understood by considering how a surfer gains energy
from the sea wave. For the wave to be damped, the wave velocity and
the surfer velocity must be similar and then the surfer is trapped by the wave.
If the surfer is slightly slower than the wave mode, the mode loses
energy to the surfer. 
A similar situation may happen in the inflating spacetime \eq{1gen-inf}.
Local fluctuations (cf., surfers) on the inflating spacetime (cf., the wave mode) are given
by Eq. \eq{1time-ncalg}. Note that these local fluctuations carry an additional localized energy
and this local energy will cause a slight delay of the drift of local lumps compared
to the inflating background. Moreover these drift delays will occur everywhere since
(quantum) fluctuations are ubiquitous. Then this is precisely the condition
for the Landau damping to occur. If this is true, the inflating mode will transfer 
its inflation (potential) energy to ubiquitous local fluctuations, 
ending the inflation through an exponential damping and
entering to a radiation dominated era via the reheating at a sufficiently high temperature
for the standard Big Bang.

The above speculation may not be so absurd,
considering the fact that the cosmic inflation is described by a conformal Hamiltonian
system \cite{libermarle,mcla-perl} which also appears in dynamical systems with friction
and the transition of such dynamical systems in nonequilibrium into equilibrium is induced
by interactions with environment. For the cosmic inflation, ubiquitous fluctuations over
the inflating spacetime will play a role of the environment. This speculation may be further 
supported by the fact that the underlying theory for emergent gravity is the Maxwell's 
electromagnetism on NC spacetime and the Landau damping can be realized even 
at a nonlinear level \cite{landau-vilani}.
Therefore it will be interesting to verify whether the naive idea can work or not.
Probably the cosmic Landau damping may be closely related to the instability of de Sitter space
suggested by Polyakov \cite{polya}.

Our real world, $\mathbb{R}^{1,3} \cong \mathbb{R} \times \mathbb{R}^{3}$, is mystic as ever
because the spatial 3-manifold does not belong to the family of (almost)
symplectic manifolds. Let us enumerate potential pathways 
leading to our tangible reality—the four-dimensional Lorentzian spacetime $\mathcal{M}$:
\begin{enumerate}
\item[A.] Analytic continuation or Wick rotation from $\mathbb{R}^4$.
\item[B.] Kaluza-Klein compactification $\mathcal{M} \times \mathbb{S}^1$.
\item[C.] Constact manifold $(\mathbb{R}^3, \eta)$.
\item[D.] Nambu structure $(\mathbb{R}^3, C)$.
\end{enumerate}
Here $\eta = dz - y dx$ is a contact form on $\mathbb{R}^3$ and $C= \frac{1}{3!} C_{\mu\nu\lambda}
dx^\mu \wedge dx^\nu \wedge dx^\lambda$ is a nondegenerate, closed three-form on $\mathbb{R}^3$.
In the case (A), the Lorentzian metric is obtained from Eq. \eq{em-metric} with $n=2$
by the Wick rotation $y^{4} = i y^0$. It is also straightforward to compactify
the $(4+1)$-dimensional Lorentzian metric \eq{eml-metric} onto $\mathbb{S}^1$ to
get the result (B). Since the time is also defined as a contact structure, the case (C)
has two contact structures as the matrix string theory discussed in Appendix C.
It may be interesting to briefly explore some clue for the cosmic inflation in the context (C).
Let $N = \mathbb{R} \times \mathbb{R}^3$ and $t \in \mathbb{R}$ be the time coordinate
and $f_t = f(t)$ be a positive monotonic function.
Define a time-dependent closed two-form on $N$ by
\begin{equation}\label{3+1}
 B_t = d\lambda_t = f_t (dT \wedge \eta + d\eta)
\end{equation}
where $\lambda_t = f_t \eta$ and $T = \ln f_t$.
Since $B_t^2 = 2 e^{2T} dT \wedge \eta \wedge d\eta$ is
nowhere vanishing, $B_t$ is a symplectic structure on $N$. Consider a time-dependent
Hamiltonian $H: N \to \mathbb{R}$ such that $dH = - e^T dT$ and
denote the Hamiltonian vector field of $H$ by $X_H$. Let $R$ be the Reeb vector field associated
with the contact form $\eta$ (see Appendix A for the definition). Then it is easy to show that
\begin{equation}\label{h-reeb}
    \iota_R B_t = dH,
\end{equation}
that is, $R = X_H$. A very interesting property is that
\begin{equation}\label{4-liouville}
    Z = \frac{\partial}{\partial T}
\end{equation}
is the Liouville vector field of the symplectic form $B_t$, i.e.,
$\mathcal{L}_Z B_t = B_t$ or $\iota_Z B_t = \lambda_t$.
This condition can be written as $\mathcal{L}_Z \lambda_t = \lambda_t$.
One can regard the Liouville vector field $Z$ as the Reeb vector field associated
with the contact form $dT$. Since $\iota_Z (B_t^2) = 2 e^{2T} \eta \wedge d\eta$,
the one-form $\lambda_t$ gives rise to a contact form on every
three-dimensional submanifold $M \subset N$ transverse to $Z$.
Thus we expect that the conformal vector field $Z$ will
generate an inflationary metric given by
\begin{equation}\label{conf-4infl}
    ds^2 = - dT^2 + e^{2T} d\mathbf{x} \cdot d\mathbf{x}.
\end{equation}
It will be interesting to have a microscopic derivation of the above inflation metric from the matrix
string theory \eq{mst}. The approach in \cite{1jun-sang} may be useful for this case.
Given our current lack of understanding in formulating emergent gravity based 
on the Nambu structure (D), the realization of this concept remains a distant aspiration.
It may be of M-theory origin because it is involved with the 3-form $C$ 
instead of symplectic 2-form $B$.

\section*{Acknowledgments}

We would like to thank Seokcheon Lee for helpful discussions and Jungjai Lee for insightful discussions on the Landau damping.
This work was supported by the National Research Foundation of Korea (NRF) with grant number NRF-2018R1D1A1B0705011314.

\appendix

\section{Locally conformal cosymplectic manifolds}

In this Appendix we briefly review locally conformal cosymplectic (LCC) manifolds. 
It was shown in \cite{clm-carolin} that an LCC manifold can be seen as a generalized
phase space of time-dependent Hamiltonian system. 
We will apply the results in Refs. \cite{vaisman2,clm-carolin} to emergent gravity and 
argue that the LCC manifold is a natural phase space describing the cosmic inflation of our universe.

First let us consider locally conformal symplectic (LCS) manifolds.
An LCS manifold is a triple $(M, \Omega, b)$ where $b$ is a closed one-form and $\Omega$ is
a nondegenerate (but not closed) two-form satisfying
\begin{equation}\label{lcs-def}
    d\Omega - b \wedge \Omega = 0.
\end{equation}
The dimension of $M$ will be assumed to be at least 4 and the one-form $b$
is called the Lee form \cite{hclee}.
A symplectic manifold corresponds to the case with $b=0$. 
If the Lee form $b$ is exact, the manifold is globally conformal symplectic (GCS).
Locally by choosing $b = d \lambda^{(\alpha)}$ for a local function $\lambda^{(\alpha)}: U_\alpha \to \mathbb{R}$
on an open neighborhood $U_\alpha$, Eq. \eq{lcs-def} is equivalent to $d(e^{-\lambda^{(\alpha)}} \Omega) = 0$,
so the local geometry of LCS manifolds is exactly the same as that of symplectic manifolds.
Thus an LCS form on a manifold $M$ is a non-degenerate two-form $\Omega$ that is locally conformal
to a symplectic form. In other words, on an LCS manifold $(M, \Omega, b)$,
there exists an open covering $\{U_\alpha\}$ of $M$ and a smooth positive function $f_\alpha$
on each $U_\alpha$
such that $f_\alpha \Omega|_{U_\alpha}$ is symplectic on $U_\alpha$. Two LCS forms $\Omega$ and $\Omega'$ are said
to be (conformally) equivalent if there exists some positive function $f$ such that
$\Omega' = f \Omega$, where the Lee form of $\Omega'$ is just $b' = b + d \ln f$.
An interesting example \cite{vaisman1} is the Hopf manifolds that
are diffeomorphic to $\mathbb{S}^1 \times \mathbb{S}^{2n-1}$ and
have a locally conformal K\"ahler metric while they admit no K\"ahler metric.

An LCS manifold can be seen as a generalized phase space of Hamiltonian dynamical systems since the form
of the Hamilton's equations is preserved by homothetic canonical transformations.
Let us recapitulate how the LCS manifolds naturally arise from the Hamiltonian dynamics of particles.
Consider a dynamical system with $n$ degrees of freedom so that its phase space is a $2n$-dimensional
differentiable manifold $M$ endowed with an open covering of coordinate neighborhoods
$\{U_\alpha\}_{\alpha  \in I}$ with local coordinates $\big( q^i_{(\alpha)}, p_i^{(\alpha)} \big),
\; i=1, \cdots, n$. Then we know that the dynamics consists of the orbits of a Hamiltonian
vector field $X_H$. Every point of $M$ has an open neighborhood $U_\alpha$ with the local Darboux
coordinates $\big( q^i_{(\alpha)}, p_i^{(\alpha)} \big)$. One can restrict the Hamiltonian $H$
and a nondegenerate two-form $\omega$ to each $U_\alpha$ to have a local Hamiltonian $H_\alpha = H_\alpha
\big( q^i_{(\alpha)}, p_i^{(\alpha)} \big)$ and a symplectic structure $\omega_\alpha = dq^i_{(\alpha)} \wedge
dp_i^{(\alpha)}$. Similarly the globally defined Hamiltonian vector field $X_H$ is restricted
to $U_\alpha$ which is precisely given by $X_{H_\alpha}$. Then the orbits are defined
by the Hamilton's equations
\begin{equation}\label{lham-eq}
    \frac{d q^i_{(\alpha)}}{dt} =  \frac{\partial H_\alpha}{\partial p_i^{(\alpha)}}, \qquad
    \frac{d p_i^{(\alpha)}}{dt} =  - \frac{\partial H_\alpha}{\partial q^i_{(\alpha)}}.
\end{equation}

When one takes the coordinate chart definition of symplectic manifolds, there is no compulsory reason
to require the two-form $\omega$ to be closed.
Indeed, the Hamiltonian formulation of particle dynamics consists in asking the local
forms $\omega_\alpha$ and local functions $H_\alpha$ to glue up to a global symplectic form $\omega$
and a global Hamiltonian $H$. However, since the dynamical information is given by a global vector field,
it is more natural to only require that the transition functions
\begin{equation}\label{trans-ftn}
q^i_{(\beta)} = q^i_{(\beta)} \big(q^i_{(\alpha)}, p_i^{(\alpha)}\big), \qquad
p_i^{(\beta)} = p_i^{(\beta)} \big(q^i_{(\alpha)}, p_i^{(\alpha)}\big)
\end{equation}
on an overlap $U_\alpha \cap U_\beta \neq \emptyset$ preserve the form of the Hamilton's
equations \eq{lham-eq}. This happens not only if Eq. \eq{trans-ftn} implies
\begin{equation}\label{gluing-symp}
 \omega_\beta = dq^i_{(\beta)} \wedge dp_i^{(\beta)} = dq^i_{(\alpha)} \wedge dp_i^{(\alpha)}
 = \omega_\alpha, \qquad H_\beta = H_\alpha,
\end{equation}
where $H_\alpha : U_\alpha \to \mathbb{R}, \; \alpha \in I$, but also if it implies
\begin{equation}\label{glue-lcs}
 \omega_\beta = \lambda_{\beta\alpha} \omega_\alpha, \qquad
 H_\beta = \lambda_{\beta\alpha} H_\alpha,
\end{equation}
where $\lambda_{\beta\alpha} = \mathrm{constant} \neq 0$. Since $\iota (X_{H_\alpha}) \omega_\alpha =
dH_\alpha$, from Eq. \eq{glue-lcs} we obtain
\begin{equation}\label{glue-lham}
 X_{H_\alpha} = X_{H_\beta},
\end{equation}
so the integral curves of $X_{H_\alpha}$ and $X_{H_\beta}$ are the same.
Furthermore, Eq. \eq{glue-lcs} implies the cocycle condition
\begin{equation}\label{cocycle}
\lambda_{\gamma\beta}  \lambda_{\beta\alpha}  = \lambda_{\gamma\alpha}
\end{equation}
as the gluing condition. We know that the cocycle condition \eq{cocycle} implies the existence of
the local functions $\sigma_\alpha: U_\alpha \to \mathbb{R}$ satisfying
\begin{equation}\label{glue-funt}
 \lambda_{\beta\alpha} = \frac{e^{\sigma_\alpha}}{e^{\sigma_\beta}}.
\end{equation}
Thus Eq. \eq{glue-lcs} shows that
\begin{equation}\label{global-lcs}
    \omega = e^{\sigma_\alpha} \omega_\alpha, \qquad H = e^{\sigma_\alpha} H_\alpha
\end{equation}
are globally defined on $M$. Moreover a Hamiltonian vector field is globally defined, i.e. $X_H = X_{H_\alpha}$,
as was indicated in Eq. \eq{glue-lham}.
Hence we have a basic line bundle $L$ over $M$ and a Hamiltonian $H$
as a cross-section of $L$ (a ``twisted Hamiltonian") instead of a simple function.
Therefore $(M, \omega)$ is an LCS manifold that can be considered as a natural phase space of
Hamiltonian dynamical systems, more general than the symplectic manifolds.

As discussed in Sec. 2, the realization of emergent geometry is intrinsically local too.
The emergent geometry is constructed by gluing local Darboux charts and their local Poisson algebras.
Therefore the construction of an LCS manifold as a generalized phase space for particle dynamics
should also be applied to the emergent geometry.
Therefore we will briefly review infinitesimal automorphisms of an LCS manifold $(M, \Omega, b)$.
The infinitesimal automorphism (IA) will be denoted by $\mathfrak{A}_\Omega$.
Let $C^\infty (M)$ denote the associative algebra of smooth functions on $M$ and $f: M \to \mathbb{R}$
be a globally defined function. The Hamiltonian vector field $X_f$ of $f \in C^\infty (M)$ with
respect to the LCS form $\Omega$ is defined by
\begin{equation}\label{ham-vec}
    \iota(X_f) \Omega = df - f b.
\end{equation}
As we observed above, there is a well-defined line bundle $L$ over $M$ in which local functions
$f_\alpha \equiv  e^{-\sigma_\alpha} f$ on a patch $U_\alpha \subset M$ correspond to sections of
$L \to U_\alpha$. If we take the Lee form on $U_\alpha$ as $b|_{U_\alpha} = d\sigma_\alpha$,
Eq. \eq{ham-vec} refers to the usual (local) Hamiltonian vector field
$X_{f_\alpha}= X_f$ defined by
\begin{equation}\label{local-hamvec}
 \iota(X_{f_\alpha}) \Omega_\alpha = df_\alpha
\end{equation}
where $\Omega_\alpha = e^{-\sigma_\alpha} \Omega$.
Using the Cartan formula for the Lie derivative
\begin{equation}\label{cartan-homo}
    \mathcal{L}_X = d\iota_X + \iota_X d,
\end{equation}
one can immediately deduce from Eqs. \eq{lcs-def} and \eq{ham-vec} that
\begin{eqnarray}\label{lie-ham1}
  && \mathcal{L}_{X_f} \Omega = b(X_f) \Omega, \\
  \label{lie-ham2}
  && \mathcal{L}_{X_f} b = d b(X_f).
\end{eqnarray}
Therefore, unlike the symplectic case, the Hamiltonian vector field $X_f$ is in general not an IA of
LCS manifolds.

Using the Hamiltonian vector fields defined by Eq. \eq{ham-vec}, we define the Poisson bracket
\begin{equation}\label{poisson-bra}
    \{f, g\}_\Omega = \iota(X_f) \iota(X_g) \Omega = - \Omega (X_f, X_g)
    = e^{\sigma_\alpha} \iota(X_{f_\alpha}) \iota(X_{g_\alpha}) \Omega_\alpha
    = e^{\sigma_\alpha} \{f_\alpha, g_\alpha \}_{\Omega_\alpha}.
\end{equation}
Then we can calculate the double Poisson bracket
\begin{equation}\label{double-bracket}
\{ \{f, g\}_\Omega, h \}_\Omega = X_h \big( \Omega (X_f, X_g) \big) - b(X_h) \Omega (X_f, X_g).
\end{equation}
Using this result, it is easy to check the Jacobi identity of the Poisson bracket:
\begin{equation}\label{jacobi-id}
 \{ \{f, g\}_\Omega, h \}_\Omega + \{ \{g, h\}_\Omega, f \}_\Omega + \{ \{h, f\}_\Omega, g \}_\Omega
 = \big( d\Omega - b \wedge \Omega \big)(X_f, X_g, X_h) = 0.
\end{equation}
Let $\mathfrak{P} = (C^\infty (M), \{-,-\}_\Omega)$ be the Poisson-Lie algebra
of $(M, \Omega)$ and $\mathfrak{X}(M)$ the Lie algebra of vector fields on $M$.
The result \eq{poisson-bra} shows that the mapping $\mathfrak{H}: \mathfrak{P} \to \mathfrak{X}(M)$ given by
$f \mapsto X_f$ is a Lie algebra homomorphism because one can derive the relation
\begin{equation}\label{plie-homo}
  X_{\{f, g\}_\Omega} = [X_f, X_g]
\end{equation}
from the Jacobi identity \eq{jacobi-id}. However, if $(M, \Omega)$ is a (connected)
LCS manifold that is not GCS, then $\mathfrak{H}$ must be a monomorphism, i.e., an injective homomorphism.
See the Proposition 2.1 in \cite{vaisman2} for the proof. This means that $X_f = 0$ implies $f=0$.
This is in stark contrast to symplectic manifolds, in which $X_f = 0$ just implies $f=\mathrm{constant}$.
Since we claim that the phase space for cosmic inflation is a locally conformal (co)symplectic manifold,
this reveals a remarkable property that vacuum energy couples to gravity and triggers cosmic inflation, as we noted before. 
However, it does not mean that the cosmological constant couples to gravity 
because physical quantities during inflation are not constant but time-dependent.

Denote the IA of $(M, \Omega)$ by $\mathfrak{X}_\Omega(M)$ whose elements obey
$\mathcal{L}_X \Omega = 0$. Then we have $\mathcal{L}_X b = 0$ by Eq. \eq{lcs-def}
which implies the condition $b(X)=\mathrm{constant}$.
In particular, if $X, Y \in \mathfrak{X}_\Omega(M)$, then $b(X)=\mathrm{constant}, \; b(Y)=\mathrm{constant}$
and $db(X,Y)=0$ yields $b([X,Y])=0$ using the formula
\begin{equation}\label{formula}
 db(X, Y) = X \big( b(Y) \big) - Y \big( b(X) \big) - b([X,Y]).
\end{equation}
Hence, the application $l: \mathfrak{X}_\Omega(M) \to \mathbb{R}$ defined by $l(X) = b(X)$ is
a Lie algebra homomorphism, called the Lee homomorphism of $\mathfrak{X}_\Omega(M)$.
The kernel $\mathrm{ker}(l)$ is the Lie algebra of the horizontal elements of $\mathfrak{X}_\Omega(M)$,
denoted by $\mathfrak{X}^{\mathrm{hor}}_\Omega(M)$. The IA $X \in \mathfrak{X}_\Omega(M)$
with $l(X) \neq 0$ is called transversal IA and an LCS manifold $M$ is called the first kind
if it has a transversal IA. Otherwise, $M$ is of the second kind and the Lee homomorphism is trivial.
Note that, if $(M, \Omega)$ is of the first kind and $f: M \to \mathbb{R}$ is a function such that
$df|_{x_0} =b(x_0)$, then $(M, e^{-f} \Omega)$ has the Lee form $b-df$ with a vanishing point,
so it becomes an LCS manifold of the second kind.

There is a special vector field $A$ defined by $\iota_A \Omega = b$. Then it is easy to see
\begin{equation}\label{char-vector}
    \iota_A b = 0, \qquad \mathcal{L}_A b = 0, \qquad \mathcal{L}_A \Omega =0.
\end{equation}
We do have $X_f \in \mathfrak{X}_\Omega(M)$ if and only if $b(X_f) = 0$ according to Eq. \eq{lie-ham1}
or equivalently $b(X_f) = \iota_{X_f} \iota_A \Omega = - \iota_A (df- fb) = - A(f)= 0$.
Let us fix an element $B \in l^{-1} (1) \subset \mathfrak{X}_\Omega(M)$. Then every element $Y$ in $\mathfrak{X}_\Omega(M)$ has a unique decomposition
\begin{equation}\label{dec-iavec}
    Y = X + l(Y) B, \qquad X \in \mathfrak{X}^{\mathrm{hor}}_\Omega(M).
\end{equation}
Now, put $a \equiv - \iota_B \Omega$, so $a(B) = 0$ and $a(A) = \iota_B \iota_A \Omega = b(B) = 1$.
Since $\mathcal{L}_B \Omega = (\iota_B d + d \iota_B)
\Omega = 0$, this yields a particular expression for $\Omega$ given by
\begin{equation}\label{omega-exp}
    \Omega = da - b \wedge a = d_b a,
\end{equation}
where $d_b$ is the Lichnerowicz differential defined by
$d_b \beta = d\beta - b \wedge \beta$ for any $k$-form $\beta$ and satisfies $d_b^2 = 0$.
Furthermore, using the formula $[\mathcal{L}_X, \iota_Y] = \iota_{[X,Y]}$ for vector fields $X$ and $Y$,
we have $\mathcal{L}_B a = 0$, hence $\iota_B da = 0$ that means rank $da < 2n$.
Since $\Omega^n \neq 0$, one can deduce from Eq. \eq{omega-exp} the condition
\begin{equation}\label{vol-ab}
  b \wedge a \wedge (da)^{n-1} \neq 0
\end{equation}
everywhere. This yields the Proposition 2.2 in Ref. \cite{vaisman2} that a manifold $M$ of dimension $2n$
admits an LCS structure of the first kind if and only if it admits two one-forms $a,b$ such that $db=0$,
rank $da < 2n$ and Eq. \eq{vol-ab} holds at every point of $M$.
Note also that $\iota_A da = \iota_A (\Omega + b \wedge a) = b - a(A) b=0$.
This means that $[A, B]=0$ because $\iota_A da = \mathcal{L}_A a = - \mathcal{L}_A \iota_B \Omega
= - \iota_{[A,B]} \Omega = 0$. In sum, there exist particular vector fields $A$ and $B$ in
$\mathfrak{X}_\Omega(M)$ that obey
\begin{equation}\label{vec-ab}
    [A, B]=0, \qquad a(A)=b(B)=1, \qquad a(B)=b(A)=0.
\end{equation}
Thus one can obtain on $M$ the vertical foliation $\mathcal{V} = \mathrm{span}\{A, B \}$,
whose leaves are the orbits of a natural action of $\mathbb{R}^2$.

Suppose that $(M, \Omega)$ is an LCS manifold of the first kind and $B$ is a basic transversal IA.
Let $\mathfrak{X}^{\mathrm{hor}}_\Omega(M, B)$ be the Lie subalgebra of
$\mathfrak{X}^{\mathrm{hor}}_\Omega(M)$ whose automorphisms also preserve $B$.
It turns out that $X \in \mathfrak{X}^{\mathrm{hor}}_\Omega(M, B)$ if and only if
$\mathcal{L}_X \Omega = 0, \; b(X) = 0$ and $[X, B]=0$.
Similarly consider the subset of $C^\infty(M)$ that consists of functions satisfying $A(f) = B(f) = 0$
and is denoted by $C_{\mathcal{V}}^\infty(M)$.
Then one can show that $\mathfrak{P}_{\mathcal{V}}= \big( C_{\mathcal{V}}^\infty(M), \{-,-\}_\Omega \big)$
is a Poisson-Lie subalgebra of $\mathfrak{P}$ and $\mathfrak{H}: \mathfrak{P}_{\mathcal{V}}
\to \mathfrak{X}^{\mathrm{hor}}_\Omega(M, B)$ is an isomorphism.
A striking fact is that a semi-simple Lie group $G$ cannot act transitively on a nonsymplectic
LCS manifold.

The formula \eq{lie-ham1} proves that a Hamiltonian vector field is a conformal infinitesimal
transformation (CIT) of $(M, \Omega)$. In general, a vector field $X$ is a CIT if
\begin{equation}\label{cit}
    \mathcal{L}_X \Omega = \alpha_X \Omega
\end{equation}
where $\alpha_X$ is a function on $M$. The CIT forms a Lie algebra denoted by $\mathfrak{X}^c_\Omega(M)$.
By differentiating Eq. \eq{cit}, one can derive that $\mathcal{L}_X b = d \alpha_X$, which implies
\begin{equation}\label{comp-rel}
    \alpha_X = b(X) + \kappa, \qquad \kappa = \mathrm{constant}.
\end{equation}
One can rewrite Eq. \eq{cit} as
\begin{equation}\label{lichner}
    \kappa \Omega = d_b (\iota_X \Omega).
\end{equation}
Thus an LCS form $\Omega$ is $d_b$-exact if there is a CIT $X$.
Or it can be written in terms of a local symplectic
form $\Omega_\alpha = e^{- \sigma_\alpha} \Omega$ as
\begin{equation}\label{cit-local}
    \mathcal{L}_{X} \Omega_\alpha = \big( \alpha_{X} - b(X) \big) \Omega_\alpha.
\end{equation}
That is, the local form of the CIT is given by
\begin{equation}\label{cit-loc}
    \mathcal{L}_{X} \Omega_\alpha = \kappa \Omega_\alpha.
\end{equation}
If we write $\Omega_\alpha = dA_{(\alpha)}$ on an open neighborhood $U_\alpha$
according to the Poincar\'e lemma, Eq. \eq{cit-loc} can be written as the form \cite{mcla-perl}
\begin{equation}\label{app-conf-vec}
    \iota_X \Omega_\alpha = \kappa A_{(\alpha)} + df_\alpha,
\end{equation}
where $f_\alpha: U_\alpha \to \mathbb{R}$ is a smooth function on $U_\alpha$.
If the conditions \eq{cit-loc} and \eq{app-conf-vec} hold either locally or globally,
we will call $X$ a conformal vector field which plays an important role in our discussion.
If $H^1 (M) = 0$, the conformal vector field $X$ has a unique decomposition given by
\begin{equation}\label{cvf-dec}
 X = \kappa Z + X_{f},
\end{equation}
where $\iota_Z \Omega =  e^{\sigma_\alpha} A_{(\alpha)} \equiv A$ 
and $\iota_{X_f} \Omega = df- fb$.
The vector field $Z$ is called the Liouville vector field \cite{libermarle}.
Note that, even though $f = 0$ identically, the conformal vector field $X = \kappa Z$
is nontrivial and it is generated by the open Wilson line \eq{1open-wilson} in our case.
We observed in Sec. 3 that this property leads to a remarkable consequence for the cosmic inflation.

We can extend the Lee homomorphism to $l: \mathfrak{X}^c_\Omega(M) \to \mathbb{R}$ by defining
$l(X) = b(X) - \alpha_X = - \kappa$. If $X, Y \in \mathfrak{X}^c_\Omega(M)$, we get
$l([X,Y])= b([X,Y])- \alpha_{[X,Y]} = - \kappa$ from $\mathcal{L}_{[X,Y]} \Omega =
\alpha_{[X,Y]} \Omega$. Hence the extended $l$ is also a Lie algebra homomorphism.
\(\)Its kernel is denoted by $\mathrm{ker}\; l = \mathfrak{X}^l_{\mathrm{Ham}}(M)$ and
consists of vector fields $X$ to obey $\mathcal{L}_X \Omega_\alpha = 0$, i.e.,
of locally Hamiltonian vector fields. Note that $\widetilde{l}(X)$ for $\widetilde{\Omega}
= e^\varphi \Omega$ is equal to $l(X)$ for $\Omega$. Thus the Lee homomorphism $l$
is conformally invariant. If we fix an element $C \in l^{-1}(1)$, we can get for every $Y \in \mathfrak{X}^c_\Omega(M)$ the unique decomposition
\begin{equation}\label{dec-cit}
    Y = X + l(Y) C, \qquad X  \in \mathfrak{X}^l_{\mathrm{Ham}}(M).
\end{equation}
Then, if $c= - \iota_C \Omega$, we can solve $\mathcal{L}_C \Omega = (\iota_C d + d \iota_C) \Omega
= \alpha_C \Omega$ to get a particular expression for $\Omega$ given by
\begin{equation}\label{omega-cit}
    \Omega = dc - b \wedge c = d_b c.
\end{equation}

In a conservative dynamical system described by a Hamiltonian vector field, time coordinate $t$ is
not a phase space coordinate but an affine parameter on particle trajectories.
But, for a general time-dependent system, it is useful to include the time coordinate as
an extra phase space coordinate. The corresponding $(2n+1)$-dimensional manifold
is known as an almost cosymplectic manifold which is a triple $(M, \Omega, \eta)$ where $\Omega$ and
$\eta$ are a two-form and a one-form on $M$ such that $\eta \wedge \Omega^n \neq 0$.
If $\Omega$ and $\eta$ are closed, i.e., $d\Omega=d\eta=0$, then $M$ is said to be
a cosymplectic manifold. Thus an odd-dimensional counterpart of a symplectic manifold is given by
a cosymplectic manifold, which is locally a product of a symplectic manifold with a circle
or a line. A contact manifold constitutes a subclass of cosymplectic manifolds with
$\Omega = d \eta$ \cite{sg-book,contact-book}. Then the one-form $\eta$ is called a contact structure or a contact one-form.
Given a contact one-form $\eta$, there is a unique vector field $R$ such that $\iota_R \eta = 1$
and $\iota_R \Omega = 0$. This vector field $R$ is known as the Reeb vector field of
the contact form $\eta$. Two contact forms $\eta$ and $\eta'$ on $M$ are equivalent
if there is a smooth positive function $\rho$ on $M$ such that $\eta' = \rho \eta$,
since $\eta' \wedge (d\eta')^n = \rho^{n+1} \eta \wedge (d\eta)^n \neq 0$.
The contact structure $C(\eta)$ determined by $\eta$ is the equivalence class of $\eta$.

The Darboux theorem for a contact manifold $(M, \eta)$ states \cite{sg-book,contact-book} that, in an open neighborhood
of each point of $M$, it is always possible to find a set
of local (Darboux) coordinates $(x^1, \cdots, x^n, y_1, \cdots, y_n, z)$ such that
the one-form $\eta$ can be written as
\begin{equation}\label{contact-form}
    \eta = dz - \sum_{i=1}^n y_i dx^i
\end{equation}
and the Reeb vector field is given by
\begin{equation}\label{reeb-vec}
    R = \frac{\partial}{\partial z}.
\end{equation}
To understand the contact one-form $\eta$ more closely,
first let us denote by $\mathcal{D}$ the contact distribution or subbundle defined by the kernel of $\eta$.
If $X, Y$ are (local) vector fields in $\mathcal{D}$, we have
\begin{equation}\label{formula-t}
 d\eta(X, Y) = X \big( \eta(Y) \big) - Y \big( \eta(X) \big) - \eta([X,Y])
 = - \eta([X,Y]).
\end{equation}
This says that the distribution is integrable if and only if $d\eta$ is zero on $\mathcal{D}$.
However the condition $\eta \wedge (d\eta)^n \neq 0$ means that the kernel of $d\eta$ is
one-dimensional and everywhere transverse to $\mathcal{D}$. Consequently, $d\eta$ is a linear symplectic
form on $\mathcal{D}$ and the largest integral submanifolds of $\mathcal{D}$ are $n$-dimensional,
so maximally non-integrable. In other words, a contact structure is nowhere integrable.
In the above Darboux coordinate system, the contact subbundle $\mathcal{D}$ is spanned by
\begin{equation}\label{span-d}
    X_i = \frac{\partial}{\partial x^i} + y_i \frac{\partial}{\partial z}, \qquad
    Y^i = \frac{\partial}{\partial y_i}, \qquad i=1, \cdots, n,
\end{equation}
so they obey the bracket relations
\begin{equation}\label{br-rel}
    [X_i, Y^j] = - \delta^j_i R, \qquad [X_i, R] = [Y^i, R] = 0.
\end{equation}
Since $d\eta = \sum_{i=1}^n dx^i \wedge dy_i$ is a symplectic form with rank $2n$,
the kernel of $d\eta$ is one-dimensional and generated by the Reeb vector $R$.
Therefore every vector field $X$ on $M$ can be uniquely written as $X = f R + Y$
where $f \in C^\infty(M)$ and $Y$ is a section of $\mathcal{D}$.
A contact structure is regular if $R$ is regular as a vector field, that is,
every point of the manifold has a neighborhood such that any integral curve of the vector field
passing through the neighborhood passes through only once.

Given a $(2n-1)$-dimensional contact manifold $M$ with a contact form $a$, i.e.
$a \wedge (da)^{n-1} \neq 0$, one can construct an LCS manifold by considering a principal
bundle $p: V \to M$ with group $\mathbb{S}^1$ over $M$.
Consider $V = \mathbb{S}^1 \times M$ endowed with the form $\Omega = da - b \wedge a = d_b a$, where
$b$ is the canonical one-form on $\mathbb{S}^1$. Clearly, $\Omega$ is nondegenerate and $b$ is
closed but not exact. And it obeys $d\Omega - b \wedge \Omega = d_b \Omega = d_b^2 a = 0$.
Hence, $(V, \Omega)$ is an LCS manifold having $b$ as its Lee form but it is not GCS.
More generally, let $p: V \to M$ be an arbitrary principal bundle with group $\mathbb{S}^1$
over a $(2n-1)$-dimensional manifold $M$. And let $a$ be the connection one-form on this
principal bundle and $F= da$ be the corresponding curvature two-form.
Then, if $b \wedge a \wedge F^{n-1} \neq 0$, the form $\Omega = F - b \wedge a$ defines an
LCS structure on $V$ which is not GCS.

Let $\mathfrak{X}(M)$ and $\Lambda^1 (M)$ be the $C^\infty (M)$-modules of differentiable
vector fields and one-forms on $M$, respectively.
If $(M, \Omega, \eta)$ is a cosymplectic manifold,
then there exists an isomorphism of $C^\infty (M)$-modules
\begin{equation}\label{module-iso}
    \Upsilon:\mathfrak{X}(M) \to \Lambda^1 (M)
\end{equation}
defined by
\begin{equation}\label{iso-def}
    \Upsilon(X) = \iota_X \Omega + \eta(X) \eta.
\end{equation}
The Reeb vector field is given by $R = \Upsilon^{-1} (\eta)$.
Let $f: M \to \mathbb{R}$ be a smooth function on $M$.
The Hamiltonian vector field $X_f$ is then defined by
\begin{equation}\label{ham-cosym}
 \Upsilon (X_f) = df - R(f) \eta + \eta.
\end{equation}
In other words, $X_f$ is the vector field characterized by the identities
\begin{equation}\label{hamvec-cos}
    \iota (X_f) \Omega = df - R(f) \eta, \qquad \eta(X_f) = 1.
\end{equation}
Then one can check that the time-like vector field $V_0$ in Eq. \eq{inner-time}
is a Hamiltonian vector field for a cosymplectic manifold $(\mathbb{R} \times M, \pi_2^* B, dt)$
where $\pi_2 : \mathbb{R} \times M \to M$ and $(M, B)$ is a symplectic manifold.

An almost cosymplectic manifold $(M, \Omega, \eta)$ is said to be LCC,
if there exist an open covering $\{U_\alpha\}_{\alpha \in I}$
and local functions $\sigma_\alpha : U_\alpha \to \mathbb{R}$ such that
\begin{equation}\label{lcc}
d(e^{-\sigma_\alpha} \Omega) = 0, \qquad d(e^{-\sigma_\alpha} \eta) = 0.
\end{equation}
The local one-forms $d \sigma_\alpha$ glue up to a closed one-form $b$ satisfying
\begin{equation}\label{lcc-def}
d\Omega - b \wedge \Omega = d_b \Omega = 0, \qquad d\eta - b \wedge \eta = d_b \eta = 0.
\end{equation}
Two LCC structures $(\Omega', \eta')$ and $(\Omega, \eta)$ are equivalent if
$\Omega' = f \Omega$ and $\eta' = f \eta$ for a positive function $f$ on $M$ where
the Lee form of $\Omega'$ is given by $b'= b + d \ln f$.
An LCC manifold reduces to a cosympletic manifold if the Lee form $b$ vanishes while
it becomes an LCS manifold if $\eta = 0$ identically.
The isomorphism \eq{iso-def} can be generalized to LCC manifolds and
the corresponding Hamiltonian vector field is defined by
\begin{equation}\label{lcc-ham}
    X_f = \Upsilon^{-1} \big(df - R(f) \eta + \eta \big) + f S
\end{equation}
where $S$ is called the canonical vector field defined by
\begin{equation}\label{lcc-can}
    \Upsilon(S) = b(R) \eta - b.
\end{equation}
Therefore, $X_f$ is characterized by the identities
\begin{equation}\label{lcc-hamvec}
    \iota(X_f) \Omega = df - R(f) \eta + f \big(b(R) \eta - b \big), \qquad
    \eta(X_f) = 1.
\end{equation}
It was shown in \cite{clm-carolin} that an LCC manifold can be seen as a generalized phase space
of time-dependent Hamiltonian systems. Hence we argue that an LCC manifold also corresponds
to a generalized phase space for an inflationary universe and its quantization realizes
a background-independent formulation of the cosmic inflation, in particular,
in the context of emergent spacetime.

\section{Harmonic oscillator with time-dependent mass}

We observed that the NC spacetime $\mathbb{R}^{2n}_\theta$ in equilibrium is described
by the Hilbert space of an $n$-dimensional harmonic oscillator while the inflating spacetime
in nonequilibrium is described by the $n$-dimensional harmonic oscillator with a negative friction.
The corresponding harmonic oscillator of constant frequency $\omega$ and friction
coefficient $\alpha$ satisfies the equation
\begin{equation}\label{inf-ho}
    \ddot{q}^i + 2 \alpha \dot{q}^i + \omega^2 q^i = 0, \qquad i=1, \cdots, n.
\end{equation}
The inflationary coordinates \eq{exp-sol} correspond to the case $\alpha = - \frac{\kappa}{2} < 0$.
It is known that the above second-order equation of motion cannot be directly derived from
the Euler-Lagrange equation of any Lagrangian.
However, there is an equivalent second-order equation
\begin{equation}\label{exp-ho}
    e^{2\alpha t} (\ddot{q}^i + 2 \alpha \dot{q}^i + \omega^2 q^i) = 0,
\end{equation}
for which a variational principle can be found \cite{d-qho}. Although Eq. \eq{inf-ho} is traditionally
considered to be non-Lagrangian, there exists an action principle for the equation of motion \eq{exp-ho}
in terms of the Lagrangian
\begin{equation}\label{exp-rlag}
    L = \frac{1}{2} m (\dot{q}^2 - \omega^2 q^2) e^{2\alpha t}.
\end{equation}
The corresponding Hamiltonian is given by
\begin{equation}\label{exp-ham}
    H = \frac{1}{2m} ( e^{-2\alpha t} p^2 +  e^{2\alpha t} m^2 \omega^2 q^2 )
\end{equation}
where $p_i = m \dot{q}^i e^{2\alpha t}$.

It is interesting to notice that the equation of motion \eq{exp-ho} can be derived from
an $n$-dimensional harmonic oscillator with a time-dependent mass $m(t)$ whose
action is given by
\begin{equation}\label{exp-act}
    S = \frac{1}{2} \int dt  \big( m (t)\dot{q}^2 - k(t) q^2 \big)
\end{equation}
where $k(t) = m(t) \omega^2$ with constant frequency $\omega$. The variational principle, $\delta S = 0$,
with respect to arbitrary variations $\delta q^i$ leads to the equation of motion
\begin{equation}\label{mass-eom}
    m(t) \Big( \ddot{q}^i +\frac{\dot{m}(t)}{m(t)}\dot{q}^i + \omega^2 q^i \Big) = 0.
\end{equation}
The second-order equation \eq{exp-ho} corresponds to the case
\begin{equation}\label{t-mass}
    \frac{\dot{m}(t)}{m(t)} = 2 \alpha \quad \Rightarrow \quad m(t) = m_0 e^{2\alpha t}.
\end{equation}
Recall that the equation of motion for the inflaton field corresponds
to the case with the time-dependent mass $m(t) = m_0 e^{3 H t}$.

There is also the first-order formalism for the dynamical system \eq{exp-act}.
The action has the form
\begin{equation}\label{1st-act}
    S = \frac{1}{2} \int dt  \big( y \dot{x} - x \dot{y} - (y^2 + 2\alpha xy + \omega^2 x^2) \big)
    e^{2\alpha t}.
\end{equation}
The equations of motion derived from the action \eq{1st-act} are given by
\begin{equation}\label{1st-eom}
    (\dot{y} + 2 \alpha y + \omega^2 x)  e^{2\alpha t} = 0,
    \qquad  (\dot{x} - y)  e^{2\alpha t} = 0.
\end{equation}
The above action \eq{1st-act} describes a singular system with second-class constraints
\begin{equation}\label{2nd-const}
    \phi_x = p_x - \frac{1}{2} y e^{2\alpha t}, \qquad \phi_y = p_y + \frac{1}{2} x e^{2\alpha t}
\end{equation}
with the Hamiltonian
\begin{equation}\label{tmass-ham}
    H(x,y, t) = \frac{1}{2}  ( y^2 + 2 \alpha x y + \omega^2 x^2) e^{2\alpha t}.
\end{equation}
Even though the constraints are explicitly time-dependent, it is still possible to apply
the Hamiltonian formalism with the help of Dirac brackets and perform the canonical quantization of
the system. It was shown in \cite{d-qho} that the classical and quantum description of the harmonic
oscillator described by the action \eq{exp-act} is equivalent to the first-order
approach given in terms of the constraint system described by the action \eq{1st-act}.
Furthermore it can be proved that the dynamical system described by Eq. \eq{exp-ho} is locally (i.e.,
$|t| < \infty$) equivalent to the system with the equation of motion \eq{inf-ho}.

\section{NC spacetime as a second-quantized string}

We know that quantum mechanics is the more fundamental description of nature than classical physics.
The microscopic world is already quantum. Nevertheless, the quantization is necessary to find
a quantum theoretical description of nature since we have understood our world starting
with the classical description which we understand better.
After quantization, the quantum theory is described by a fundamental NC algebra
such as Eq. \eq{nc-phase}. A striking feature of the NC algebra $\mathcal{A}_\hbar$ is that
every point in $\mathbb{R}^n$ is unitarily
equivalent because translations in $\mathbb{R}^n$ are generated by an inner automorphism of
$\mathcal{A}_\hbar$, i.e., $f(x+a) = U(a) f(x) U(a)^\dagger$ where $f(x) \in \mathcal{A}_\hbar$
and $U(a) = e^{ip_i a^i/\hbar} \in \mathrm{Inn}(\mathcal{A}_\hbar)$.
Therefore, through the quantization, the concept of (phase) space is doomed.
Instead the (phase) space is replaced by the algebra $\mathcal{A}_\hbar$ and
its Hilbert space representation and dynamical variables become operators acting on the Hilbert space.
Only in the classical limit, a phase space with the symplectic structure $\omega =  dx^i \wedge dp_i$
is emergent from the quantum algebra $\mathcal{A}_\hbar$ such as \eq{nc-phase}.

Recall that the mathematical structure of NC spacetime is basically the same as
the NC phase space in quantum mechanics \cite{hsy-rmpla}.
Therefore essential features in quantum mechanics must be
applied to the NC spacetime too. In particular, NC algebras $\mathcal{A}_\theta$
such as the NC space \eq{1extra-nc2n} also play
a fundamental role and every points in the NC space are indistinguishable, i.e.,
unitarily equivalent because any two points are connected by an inner automorphism of $\mathcal{A}_\theta$.
This implies that there is no concept of space(time) in the NC algebra $\mathcal{A}_\theta$
for the same reason as quantum mechanics and
a classical spacetime must be derived from the NC algebra $\mathcal{A}_\theta$.
After all, an important lesson is that NC spacetime necessarily implies emergent spacetime.

Although spacetime at a microscopic scale,  e.g. the Planck scale $L_P$, is intrinsically NC, we understand
the NC spacetime through the quantization of a symplectic (or more generally Poisson) manifold.
Let $(M, B)$ be a symplectic manifold. On the one hand, the basic concept in symplectic geometry is an area
defined by the symplectic two-form $B$ that is a nondegenerate, closed two-form.
On the other hand, the basic concept in Riemannian geometry determined by a pair $(M, g)$ is
a distance defined by the metric tensor $g$ that is a nondegenerate, symmetric bilinear form.
One may identify this distance with a geodesic worldline of a ``particle" moving in $M$.
Geodesic curves in $M$ give us all information of Riemannian geometry $(M, g)$.
On the contrary, the area in symplectic geometry $(M, B)$ may be regarded as a minimal worldsheet
swept by a ``string" moving in $M$. In this picture, the wiggly string, so a fluctuating worlsheet,
corresponds to a deformation of symplectic structure in $M$. This picture becomes more transparent
by the so-called pseudoholomorphic or $J$-holomorphic curve introduced by Gromov \cite{gromov}.

Let $(M, J)$ be an almost complex manifold and $(\Sigma, j)$ be a Riemann surface.
By the compatibility of $J$ to $B$, we have the relation $g(X, Y) = B(X, JY)$
for any vector fields $X, Y \in \mathfrak{X}(M)$. Let us also fix a Hermitian metric $h$ of $(\Sigma, j)$.
A smooth map $f:\Sigma \to M$ is called pseudoholomorphic \cite{donaldson}
if the differential $df: T\Sigma \to TM$
is a complex linear map with respect to $j$ and $J$:
\begin{equation}\label{holo-curve}
    df \circ j = J \circ df.
\end{equation}
This condition corresponds to the commutativity of the following diagram
\begin{displaymath}\label{j-holo}
     \xymatrix{
         T\Sigma \ar[r]^j \ar[d]_{df} & T\Sigma \ar[d]^{df} \\
         TM \ar[r]_{J}       & TM }
\end{displaymath}
Since $J^{-1} = - J$, it is also equivalent to $\overline{\partial}_J f = 0$
where $\overline{\partial}_J f := \frac{1}{2} (df + J \circ df \circ j)$.
For example, suppose that the Riemann surface is $(\Sigma, i)$ where $i$ is
the standard complex structure. We can work in a chart $u_\epsilon : U_\epsilon \to \mathbb{C}$
with local coordinate $z = \tau + i \sigma$ where $U_\epsilon \subset \Sigma$ is an open neighborhood.
Define $f_\epsilon = f \circ u_\epsilon^{-1}$. In this case, we have
\begin{equation}\label{curve-h}
\overline{\partial}_J f = \frac{1}{2} \left[ \Big(\frac{\partial f_\epsilon}{\partial \tau}
+ J(f_\epsilon) \frac{\partial f_\epsilon}{\partial \sigma}\Big) d\tau
+ \Big(\frac{\partial f_\epsilon}{\partial \sigma}
- J(f_\epsilon) \frac{\partial f_\epsilon}{\partial \tau}\Big) d\sigma \right].
\end{equation}
Thus we see that $\overline{\partial}_J f = 0$ if
\begin{equation}\label{curve-heq}
 \frac{\partial f_\epsilon}{\partial \tau}
+ J(f_\epsilon) \frac{\partial f_\epsilon}{\partial \sigma} = 0.
\end{equation}

Since $J$ is $B$-compatible, every smooth map $f:\Sigma \to M$ satisfies \cite{duff-sala,holo-review}
\begin{equation}\label{j-curve}
    \frac{1}{2} \int_\Sigma ||df||_g^2 \; d\mathrm{vol}_\Sigma
    =  \int_\Sigma ||\overline{\partial}_J f||_g^2 \; d\mathrm{vol}_\Sigma +  \int_\Sigma f^* B,
\end{equation}
where the norms are taken with respect to the metric $g$ and $d\mathrm{vol}_\Sigma$ is
a volume form on $\Sigma$. In terms of local coordinates, $(\sigma^1, \sigma^2)$ on $\Sigma$
and $f(\sigma) = (x^1, \cdots, x^{2n})$ on $M$,
\begin{equation}\label{dudu}
    ||df||_g^2 = g_{\mu\nu} \big( f(\sigma) \big) \frac{\partial x^\mu}{\partial \sigma^\alpha}
    \frac{\partial x^\nu}{\partial \sigma^\beta} h^{\alpha\beta} (\sigma) 
\end{equation}
and $d\mathrm{vol}_\Sigma = \sqrt{h} d^2 \sigma$.
Therefore, the left-hand side of Eq. \eq{j-curve} is nothing but the Polyakov action
in string theory. For a pseudoholomorphic curve $f:\Sigma \to M$ that obeys $\overline{\partial}_J f = 0$,
we thus have the identity
\begin{equation}\label{j-area}
   S_P (f) \equiv \frac{1}{2} \int_\Sigma ||df||_g^2 \; d\mathrm{vol}_\Sigma = \int_\Sigma f^* B.
\end{equation}
This means that any pseudoholomorphic curves minimize the ``harmonic energy" $S_P (f)$
in a fixed homology class and so are harmonic maps. In other words, their symplectic area coincides with
the surface area. Therefore, any pseudoholomorphic curve is a solution of
the worldsheet Polyakov action $S_P (f)$.
For instance, if $M = \mathbb{C}^n$ with complex coordinates $\phi^i = x^{2i-1} + \sqrt{-1} x^{2i} \;
(i=1, \dots, n)$ and $f_\epsilon (z, \bar{z}) \equiv \phi^i (z, \bar{z})$, Eq. \eq{curve-heq} becomes
\begin{equation}\label{cr-curve}
 \frac{1}{2} \Big( \frac{\partial}{\partial \tau}
+ \sqrt{-1} \frac{\partial}{\partial \sigma} \Big) \phi^i (z, \bar{z})
= \partial_{\bar{z}} \phi^i (z, \bar{z}) = 0.
\end{equation}
In this case, pseudoholomorphic curves coincide with holomorphic curves.
Moreover such curves are harmonic and minimal surfaces.\footnote{In the topological A-model that
is concerned with pseudoholomorphic maps from $\Sigma$ to $M = T^* N$,
there is a vanishing theorem \cite{witten-csst} stating that $\int_\Sigma f^* B = 0$.
In particular, the mappings from $\partial \Sigma$ to $N$ are necessarily constant.}

The pseudoholomorphic curve also provides us a useful tool to understand the emergent gravity picture.
To demonstrate this aspect, let us include a boundary interaction in the sigma model \eq{j-curve}
such that the open string action is given by
\begin{equation}\label{open-poly}
   S_A (f) \equiv \frac{1}{2} \int_\Sigma ||df||_g^2 \; d\mathrm{vol}_\Sigma
   + \int_{\partial \Sigma} f^* A,
\end{equation}
where the one-form $A$ is the connection of a line bundle $L \to M$.
Using the Stokes' theorem, the second term can be written as
\begin{equation}\label{stokes}
\int_{\partial \Sigma} f^* A = \int_{\Sigma} f^* dA.
\end{equation}
After combining the identities \eq{j-curve} and \eq{stokes} together, we write the action
\begin{equation}\label{open-curve}
   S_A (f) = \int_\Sigma ||\overline{\partial}_J f||_g^2 \; d\mathrm{vol}_\Sigma
   +  \int_\Sigma f^* \mathcal{F},
\end{equation}
where $\mathcal{F} = B + F$ and $F= dA$. If one recalls the derivation of Eq. \eq{j-curve},
one may immediately realize that the action $S_A (f)$ can equivalently be written as the form
of the Polyakov action
\begin{equation}\label{open-holo}
S_P (\psi) \equiv \frac{1}{2} \int_\Sigma ||d\psi||_{\mathcal{G}}^2 \; d\mathrm{vol}_\Sigma,
\end{equation}
where the differential $d\psi$ for a smooth map $\psi: \Sigma \to M$ has the norm taken with respect to
some metric $\mathcal{G}$. For this purpose, let us assume that the almost complex structure $J$ is
also compatible with the deformed symplectic structure $\mathcal{F}$, i.e.,
\begin{equation}\label{j-compa}
    \mathcal{G}(X, Y) = \mathcal{F}(X, JY), \qquad \forall X, Y \in \mathfrak{X}(M)
\end{equation}
is a Riemannian metric on $M$.
An explicit representation of the Polyakov action \eq{open-holo} can be made by introducing
local coordinates $\psi(\sigma) = (X^1, \cdots, X^{2n})$ on an open set $U_i \subset M$ so that
\begin{equation}\label{dpdp}
    ||d\psi||_{\mathcal{G}}^2 =   {\mathcal{G}}_{\mu\nu} \big( \psi(\sigma) \big)
    \frac{\partial X^\mu}{\partial \sigma^\alpha}
    \frac{\partial X^\nu}{\partial \sigma^\beta} h^{\alpha\beta} (\sigma).
\end{equation}
One can then apply the same derivation of Eq. \eq{j-curve} to the action \eq{open-holo} to derive
the identity
\begin{equation}\label{open-id}
   \frac{1}{2} \int_\Sigma ||d\psi||_{\mathcal{G}}^2 \; d\mathrm{vol}_\Sigma
   = \int_\Sigma ||\overline{\partial}_J \psi||_{\mathcal{G}}^2 \; d\mathrm{vol}_\Sigma
   +  \int_\Sigma \psi^* \mathcal{F}.
\end{equation}
For pseudoholomorphic curves $\psi: \Sigma \to M$ satisfying $\overline{\partial}_J \psi = 0$,
we finally get the result
\begin{equation}\label{open-area}
   S_P (\psi) = \frac{1}{2} \int_\Sigma ||d\psi||_{\mathcal{G}}^2 \; d\mathrm{vol}_\Sigma
   =  \int_\Sigma \psi^* \mathcal{F}.
\end{equation}

The above argument reveals a nice picture that dynamical $U(1)$ gauge fields in a line bundle
$L$ over $M$ deform an underlying symplectic structure $(M, B)$ and
this deformation is transformed into the dynamics of gravity \cite{1q-emg}. 
As we observed before, the symplectic geometry is probed by strings while
the Riemannian geometry is probed by particles.
We note that the NC space \eq{1extra-nc2n} defines only a minimal area $\alpha' = l_s^2$
whereas the concept of point is doomed as if $\hbar$ in quantum mechanics introduces a minimal area
in the NC phase space \eq{nc-phase}. The minimal area (surface) in the NC space
behaves like the smallest unit of spacetime blob and acts as a basic building block of string theory.
The concept of pseudoholomorphic or $J$-holomorphic curves in symplectic geometry
plays a role of such minimal surfaces.
It is known \cite{donaldson} that there is a nonlinear Fredholm theory which describes
the deformations of a given pseudoholomorphic curve $f: \Sigma \to (M, J)$ and the deformations
are parameterized by a finite-dimensional moduli space. (This moduli space may be enriched by
considering pseudoholomorphic curves in an LCS manifold.)
When a symplectic manifold is probed with a string or pseudoholomorphic curve,
the notion of a wiggly string in this probe picture corresponds to the deformation
of a symplectic structure. Hence the emergence of gravity from symplectic geometry or
more precisely NC $U(1)$ gauge fields may be reasonable because we know from string theory
that a Riemannian geometry (or general relativity) is emergent from the wiggly string.

We can think of the integral $A(f) = \int_\Sigma f^* B$ in two ways if $f$ is a pseudoholomorphic curve.
On the one hand, the pointwise compatibility between the structures $(B, J)$ means that
$A(f)$ is essentially the area of the image of $f$, measured in the Riemannian metric $g$.
On the other hand, the condition that $B$ is closed means that $A(f)$ is a topological (homotopy)
invariant of the map $f$ since it depends only on the evaluation of a closed 2-form $B$ on the 2-chain
defined by $f(\Sigma)$. Hence we can use the curves in two main ways \cite{donaldson}.
The first way is as geometrical probes to explore a symplectic manifold, as we advocated above.
The second way is as the source of numerical invariants known as the Gromov-Witten invariants.
Using the pseudoholomorphic curves, Gromov proved a surprising non-squeezing
theorem \cite{gromov,duff-sala,holo-review} stating that a ball $B_{2n}(r)$ of radius $r$
in a symplectic vector space $\mathbb{R}^{2n}$ with the standard symplectic form $B$
cannot be mapped by a symplectomorphism into any cylinder $B_2 (R) \times \mathbb{R}^{2n-2}$
of radius $R$ if $R  < r$. It is possible to replace $\mathbb{R}^{2n-2}$ by a $(2n-2)$-dimensional
compact symplectic manifold $V$ with $\pi_2 (V) = 0$.

Now we will discuss how a NC space provides us an important clue for a background-independent
formulation of string theory. The NC spacetime is defined by the quantization of a symplectic manifold $(M, B)$.
One may try to lift the notion of the pseudoholomorphic curve to a quantized symplectic manifold,
namely, a NC space such as Eq. \eq{1extra-nc2n}. The quantization of a symplectic manifold leads to
a radical change of classical concepts such as spaces and observables. The classical space is
replaced by a Hilbert space and dynamical observables become operators acting on the Hilbert space.
Then, as we discussed in Sec. 2, the NC spacetime will provide a more elegant framework
for the background-independent formulation of quantum gravity in terms of matrix models,
which is still elusive in string theory. Recall that the dynamical Lorentzian
spacetime \eq{eml-metric} emerges from a classical solution of the matrix model \eq{bfss},
and the cosmic inflation described by the metric \eq{1gen-inf} also arises
as a solution of the time-dependent matrix model.

In order to grasp how a pseudoholomorphic curve looks like in NC spacetime,
let us consider the simplest case in Eq. \eq{cr-curve}.
After quantization, the coordinates of $\mathbb{C}^n$ denoted by $\phi^i (z, \bar{z})$ become operators
in a NC $\star$-algebra $\mathcal{A}_\theta^2 \equiv \mathcal{A}_\theta
\big( C^\infty (\mathbb{R}^{2}) \big) = C^\infty (\mathbb{R}^{2}) \otimes \mathcal{A}_\theta$,
i.e., $\phi^i (z, \bar{z}) \to \widehat{\phi}^i (z, \bar{z}) \in \mathcal{A}_\theta^2$.
The worldsheet $\mathbb{R}^2$ may be replaced by $\mathbb{T}^2, \; \mathbb{R} \times \mathbb{S}^1$ or $\mathbb{S}^2$.
Let us clarify the notation $\mathcal{A}_\theta^2$ after the Wick rotation of the worldsheet
coordinate $\tau = i t$, so $\mathbb{R}^2 \to \mathbb{R}^{1,1}$.
Consider a generic element in the NC $\star$-algebra $\mathcal{A}_\theta^2$ given by
\begin{equation}\label{nc-2alg}
    \widehat{f} (t, \sigma, y) \in \mathcal{A}_\theta^2.
\end{equation}
The matrix representation \eq{mat-time} is now generalized to
\begin{equation}\label{mat-2time}
    \widehat{f}(t, \sigma, y) = \sum_{n,m=1}^\infty |n\rangle \langle n |\widehat{f}(t, \sigma, y)
    |m \rangle \langle m|
    = \sum_{n,m=1}^\infty f_{nm} (t, \sigma) |n \rangle \langle m|
\end{equation}
where the coefficients $f_{nm} (t, \sigma):= [f(t, \sigma)]_{nm}$ are elements of
a matrix $f(t, \sigma)$ in $\mathcal{A}^2_N \equiv \mathcal{A}_N \big( C^\infty (\mathbb{R}^{1,1}) \big)
= C^\infty (\mathbb{R}^{1,1}) \otimes \mathcal{A}_N$ as a representation of the observable \eq{nc-2alg}
on the Hilbert space \eq{hilbert}.
Then we have an obvious generalization of the duality chain \eq{tduality-chain} as follows:
\begin{equation}\label{2duality-chain}
  \mathcal{A}^2_N  \quad \Longrightarrow \quad \mathcal{A}^2_\theta \quad
  \Longrightarrow \quad \mathfrak{D}^2.
\end{equation}
The module of derivations is similarly a direct sum of the submodules of horizontal
and inner derivations \cite{azam}:
\begin{equation}\label{2deriv}
  \mathfrak{D}^2 = \mathrm{Hor}(\mathcal{A}^2_N) \oplus \mathfrak{D} (\mathcal{A}^2_N) \cong
  \mathrm{Hor}(\mathcal{A}^2_\theta) \oplus \mathfrak{D} (\mathcal{A}^2_\theta),
\end{equation}
where horizontal derivations are locally generated by a vector field
\begin{equation}\label{2time-vec}
    k(t, \sigma, y) \frac{\partial}{\partial t} + l(t, \sigma, y) \frac{\partial}{\partial \sigma}
    \in \mathrm{Hor}(\mathcal{A}^2_\theta).
\end{equation}

It can be shown \cite{1hsy-review,1q-emg} that the matrix model for the duality chain \eq{2duality-chain}
is given by
\begin{equation}\label{mst}
    S = - \frac{1}{g_s^2} \int d^2 \sigma \Tr \Big( \frac{1}{4} F_{\alpha\beta}^2
    + \frac{1}{2}  (D_\alpha \phi_a)^2 - \frac{1}{4}[\phi_a, \phi_b]^2 \Big),
\end{equation}
where $a = 2, \cdots, 2n+1$ and $\sigma^\alpha = (t, \sigma), \; \alpha = 0, 1$ and
$F_{\alpha\beta} = \partial_\alpha A_\beta - \partial_\beta A_\alpha - i [A_\alpha, A_\beta]$.
The $n = 4$ case is known as the matrix string theory that is supposed to describe a nonperturbative
type IIA string theory in light-cone gauge \cite{dvv}. 
The matrix string theory can also be obtained from
the BFSS matrix model via compactification on a circle \cite{1wtaylor}.
To achieve this model, the BFSS matrix model has to have 9 adjoint scalar fields,
$\phi_a (t) \; (a=1, \cdots, 9)$, unlike the action \eq{bfss} with even number of adjoint scalar fields.
The equivalence \eq{bfss-u1} can be realized only in the case of even number of adjoint scalar fields. In this case, the action \eq{bfss} can be understood as
a Hilbert space representation of certain NC gauge theory under a symplectic vacuum such as
\eq{moyal-vac} with rank$(B)= 2n$. However we do not know a corresponding NC gauge theory whose
Hilbert space representation precisely reproduces the BFSS matrix model.
Fortunately the matrix string theory \eq{mst} has
8 adjoint scalar fields for $n=4$. Thus it is possible to realize it as the Hilbert space
representation of $(9+1)$-dimensional NC $U(1)$ gauge theory
with rank$(B)= 8$.

It will be interesting to understand how to derive the matrix string
theory \eq{mst} from the MQM \eq{bfss} as if the latter has been derived
from a contact structure of the zero-dimensional matrix model \eq{ikkt}.
The basic idea is similar to the scheme to construct the one-dimensional matrix model \eq{bfss}
through the contact structure of zero-dimensional matrices. A difference is that we start with
the one-dimensional matrix model \eq{bfss} and introduce an additional contact structure along
a spatial direction whose coordinate is called $\sigma$ in our case.
Ultimately, the matrix string theory \eq{mst} can be realized as the quantization of a regular
2-contact manifold. See Ref. \cite{vaisman2} for a general $k$-contact manifold.
First let us consider the projection $\pi_2: \mathbb{R}^{1,1} \times M \to M, \; \pi_2(\sigma^\alpha, x) = x$
where $M$ is a symplectic manifold with the symplectic form $B$.\footnote{It is possible to replace
$\mathbb{R}^{1,1} \times M$ by a general $(2n+2)$-dimensional manifold $N$ as far as there is a
well-defined two-dimensional foliation $\mathcal{V}$ such that the corresponding space of leaves
$N/\mathcal{V} = M$ is a Hausdorff differentiable manifold \cite{vaisman2}. See \eq{vec-ab}
for a relevant discussion. We will keep the maximal simplicity for a plain argument.}
The regular 2-contact $(2n+2)$-dimensional manifold is defined by a quartet $(\mathbb{R}^{1,1} \times M,
\widetilde{B}, \eta^\alpha), \; \alpha = 0, 1$, where $\widetilde{B} = \pi_2^* B$, such that
\begin{equation}\label{2-contact}
    \eta^0 \wedge \eta^1 \wedge B^n \neq 0
\end{equation}
everywhere and $d\eta^\alpha = \gamma^\alpha B$ with constants $\gamma^\alpha$ and $d B = 0$.
Moreover there are uniquely defined two Reeb vectors $R_\alpha \; (\alpha = 0, 1)$ satisfying
\begin{equation}\label{2-reeb}
    \iota_{R_\alpha} \eta^\beta = \delta^\beta_\alpha, \qquad \iota_{R_\alpha} B = 0,
    \qquad \alpha, \beta = 0, 1.
\end{equation}
The above relations imply
\begin{equation}\label{2-cont-id}
    \mathcal{L}_{R_\alpha} \eta^\beta = 0, \qquad \mathcal{L}_{R_\alpha} B = 0, \qquad [R_0, R_1] = 0.
\end{equation}
For example, the contact forms for the matrix string theory \eq{mst} are given by
\begin{equation}\label{mst-cform}
    \eta^0 = dt - \frac{1}{2} p_a dy^a, \qquad \eta^1 = d \sigma - \frac{1}{2} p_a dy^a,
\end{equation}
which determines the corresponding Reeb vectors
\begin{equation}\label{mst-reeb}
    R_0 = \frac{\partial}{\partial t}, \qquad  R_1 = \frac{\partial}{\partial \sigma}.
\end{equation}
These Reeb vectors span the space of horizontal derivations in Eq. \eq{2time-vec}.

Since there are two independent contact structures, each contact structure generates its own
Hamiltonian vector field defined by \eq{hamvec-cos}. For the contact structures in Eq. \eq{mst-cform},
they are given by
\begin{equation}\label{mst-hvec}
    V_\alpha = \frac{\partial}{\partial \sigma^\alpha} + A_\alpha^\mu (t, \sigma, y)
    \frac{\partial}{\partial y^\mu}.
\end{equation}
The quantization of the 2-contact manifold $(\mathbb{R}^{1,1} \times M, \widetilde{B}, \eta^\alpha)$ is
simple because it is performed using the Darboux coordinates $(\sigma^\alpha, y^a)$.
It is basically defined by the quantization of the symplectic manifold $(M, B)$ in which $\sigma^\alpha$
are regarded as classical variables like the time coordinate in the algebra $\mathcal{A}_\theta^1$.
After quantization, a generic element of the NC $\star$-algebra $\mathcal{A}_\theta^2$ takes
the form \eq{nc-2alg}. Then the module $\mathfrak{D}^2$ in Eq. \eq{2deriv} is generated by
\begin{equation}\label{mst-derivation}
\mathfrak{D}^2 = \Big\{ \widehat{V}_A (t, \sigma) = \big( \widehat{V}_\alpha,
\widehat{V}_a \big) (t, \sigma) | \widehat{V}_\alpha (t, \sigma) = \frac{\partial}{\partial \sigma^\alpha} + \mathrm{ad}_{\widehat{A}_\alpha}, \;
\widehat{V}_a (t, \sigma) = \mathrm{ad}_{\widehat{\phi}_a} \Big\},
\end{equation}
where $A=0, 1, \cdots, 2n+1$ and the adjoint operations are inner derivations of $\mathcal{A}_\theta^2$.
In the commutative limit, the module \eq{mst-derivation} reduces to ordinary vector fields 
$V_A = (V_\alpha, V_a) \in \mathfrak{X}(\mathcal{M})$ and it is related to the orthonormal 
frames by $V_A = \lambda E_A$.
Finally the corresponding Lorentzian metric dual to the matrix string
theory \eq{mst} is given by \cite{1hsy-review,1q-emg}
\begin{equation}\label{mst-gmetric}
    ds^2 = \lambda^2 \eta_{AB} v^A \otimes v^B =
    \lambda^2 \big( \eta_{\alpha\beta} d\sigma^\alpha d \sigma^\beta
    + v^a_\mu v^a_\nu (dy^\mu - \mathbf{A}^\mu) (dy^\nu - \mathbf{A}^\nu) \big),
\end{equation}
where $\mathbf{A}^\mu := A_\alpha^\mu (t, \sigma, y) d \sigma^\alpha$ and
$\lambda^2 = \nu_{(t, \sigma)} (V_0, V_1, \cdots, V_{2n+1})$ is determined by the volume preserving
condition, $\mathcal{L}_{V_A} \nu_{(t, \sigma)} = 0$, with respect to a given volume form
\begin{equation}\label{mst-vol}
\nu_{(t, \sigma)} = dt \wedge d \sigma \wedge \nu = \lambda^2  dt \wedge d \sigma \wedge
v^1 \wedge \cdots \wedge v^{2n}.
\end{equation}

Instead of the conformal frame $V_A = \lambda E_A$, one may choose another frame, 
the so-called comoving frame, similar to Eq. \eq{lorentz-iviel}:
\begin{equation}\label{mst-comframe}
V_A = (V_\alpha, V_a) = (E_\alpha, \lambda E_a).
\end{equation}
The $(2n+2)$-dimensional Lorentzian metric is then given by 
\begin{equation}\label{mst-commetric}
    ds^2 = \eta_{AB} e^A \otimes e^B =
    \eta_{\alpha\beta} d\sigma^\alpha d \sigma^\beta
    +  \lambda^2 v^a_\mu v^a_\nu (dy^\mu - \mathbf{A}^\mu) 
    (dy^\nu - \mathbf{A}^\nu). 
\end{equation}
This comoving frame may be more convenient to incorporate the inflation metric 
\eq{conf-4infl}.

Let us come back to our previous question about the generalization of pseudoholomorphic curves. In order to address this issue, let us consider the Wick rotation $t = - i \tau$
again to return to the Euclidean space. If the quantum version of pseudoholomorphic curves exists,
Eq. \eq{curve-heq} suggests that it will also obey the first-order partial differential equations.
It is well-known \cite{mst-bps} that the matrix string action \eq{mst} admits such a first-order system.
For simplicity, assume that adjoint scalar fields mostly vanish except $(\phi_2, \phi_3) \neq 0$.
It is convenient to use the complex variables
\begin{equation}\label{cpx-adj}
    \phi = \frac{1}{2} (\phi_2 - i \phi_3), \qquad \phi^\dagger = \frac{1}{2} (\phi_2 + i \phi_3).
\end{equation}
It is not difficult to show that the Euclidean action with $\phi_a = 0$ for $a=4, \cdots, 9$ can be
written as the Bogomol'nyi-type, i.e.,
\begin{eqnarray}\label{mst-bogomol}
    S &=& \frac{1}{g_s^2} \int d^2 \sigma \Tr \Big( \frac{1}{4} F_{\alpha\beta}^2
    + \frac{1}{2}  (D_\alpha \phi_a)^2 - \frac{1}{4}[\phi_a, \phi_b]^2 \Big) \nonumber \\
    &=& \frac{2}{g_s^2} \int d^2 \sigma \Tr \Big(  \big( i F_{z\bar{z}} - [\phi, \phi^\dagger] \big)^2
    + |D_{\bar{z}} \phi |^2 - i \partial_\alpha \big( \varepsilon^{\alpha\beta}
    \phi^\dagger D_\beta \phi \big) \Big).
\end{eqnarray}
Since the last term is a topological number, the minimum of the action is achieved in
the configurations obeying
\begin{equation}\label{hitchin}
   F_{z\bar{z}} + i [\phi, \phi^\dagger] = 0, \qquad D_{\bar{z}} \phi =0.
\end{equation}
Note that the above equations recover Eq. \eq{cr-curve} in a very commutative limit
where $[\phi^\dagger, \phi] = 0$. Therefore it is reasonable to identify Eq. \eq{hitchin} with
the quantum version of pseudoholomorphic curves.

Mathematically Eq. \eq{hitchin} is equivalent to the Hitchin equations describing
a Higgs bundle \cite{hitchin}. A Higgs bundle is a system composed of a connection $A$
on a principal $G$-bundle or simply a vector bundle $E$ over a Riemann surface $\Sigma$
and a holomorphic endomorphism $\phi$ of $E$ satisfying Eq. \eq{hitchin}.
The Hitchin equations describe four-dimensional Yang-Mills instantons on $\Sigma \times \mathbb{R}^2$
which are invariant with respect to the translation group $\mathbb{R}^2$. (This $\mathbb{R}^2$ is
transverse to the Riemann surface.)
Using the translation invariance, the Yang-Mills instantons can be dimensionally reduced
to the Riemann surface $\Sigma$ in which Yang-Mills gauge fields along the isometry directions
become an adjoint Higgs field $\phi$. In our case the gauge group $G$ is $U(N)$. In particular,
we are interested in the large $N$ limit, i.e., $N \to \infty$.
In this limit, the action \eq{mst-bogomol} can be mapped to four-dimensional NC $U(1)$ gauge
theory under the Coulomb branch vacuum $\langle \phi_a \rangle_{\mathrm{vac}} = p_a, \; a= 2, 3$
obeying the commutation relation $[p_2, p_3] = -i B_{23}$. Then the Hitchin equations \eq{hitchin}
precisely become the self-duality equation for NC $U(1)$ instantons on $\Sigma 
\; (\mathrm{or} \; \mathbb{R}^2) \times \mathbb{R}^2_\theta$ \cite{nc-inst,kly-plb}.
The corresponding gravitational metric for the case $n=1$ can be
identified with Eq. \eq{mst-gmetric} with the analytic continuation $t =-i\tau$.
It was shown in \cite{1hsy-epjc09,1lry-jhep,hsy-epl} that the solution of the Hitchin
equations \eq{hitchin} is dual to four-dimensional gravitational instantons
which are hyper-K\"ahler manifolds. In particular, the real heaven is governed by
the $su(\infty)$ Toda equation and the self-duality equation for the real heaven
exactly reduces to the commutative limit of the Hitchin equations \eq{hitchin}.
See eq. (4.31) in Ref. \cite{1lry-jhep}. Thus the Hitchin system with the gauge group
$G = U(N \to \infty)$ may be closely related to the Toda field theory.
Indeed this interesting connection was already analyzed in \cite{h-toda}.
In sum, Hitchin equations, NC $U(1)$ instantons, gravitational instantons and pseudoholomorphic
curves may be only the tip of the iceberg in the matrix string theory \eq{mst} that have barely
shown themselves.

Let us conclude this section by drawing an invaluable insight. 
We have observed that NC spacetime is much more radical and mysterious than we
thought before. It is fair to say that we have not yet fully understood 
the mathematical foundation of NC spacetime. 
A remarkable point is that NC spacetime necessarily implies emergent spacetime if spacetime at microscopic scales should be viewed as NC. 
This means that classical spacetime is a derived concept from something deeper. 
A pseudoholomorphic curve is a stringy generalization of a geodesic worldline in Riemannian geometry \cite{donaldson}. 
Recall that the pseudoholomorphic curve is basically a minimal surface or a string worldsheet embedded into spacetime. 
However, to make sense of the emergent spacetime picture, 
we need a mathematically precise framework for describing
strings in a background-independent way. The background-independent theory 
must give up the picture that strings are vibrating in a preexisting spacetime. 
In this Appendix, we have aimed at clarifying how the pseudoholomorphic curves can be 
lifted to a NC spacetime by the matrix string theory. The matrix string theory naturally extends the first-quantized string theory 
so that it also describes the non-perturbative interactions of splitting and joining of strings, producing surfaces with nontrivial topology \cite{dvv}. 
That is, the matrix string theory is a second-quantized theory in
which spacetime emerges from the collective behavior of matrix strings. 
Thus we argue that the NC spacetime can be viewed as a second-quantized string 
from the perspective of the background-independent formulation of quantum gravity.

%\newpage

\end{document}